\documentclass[fleqn,usenatbib]{mnras}
\hypersetup{linkcolor=red}          
\usepackage{multirow}
\usepackage[dvipsnames]{xcolor}
\usepackage{newtxtext,newtxmath}
\usepackage{orcidlink}

\def\equationautorefname~#1\null{Eq.~#1\null}
 
\defcitealias{Zhu_et_al.(2023a)}{Paper I}
\defcitealias{Lu_et_al.(2023a)}{Paper II}
\defcitealias{Zhu_et_al.(2024)}{Paper III}
   
\title[DM and IMF in MaNGA]{MaNGA DynPop -- V. The dark-matter fraction versus stellar velocity dispersion relation and stellar initial mass function variations in galaxies: dynamical models and full spectrum fitting of integral-field spectroscopy}

\author[S. Lu et al.]
{Shengdong Lu\orcidlink{0000-0002-6726-9499}$^{1}$\thanks{E-mail: \url{lushengdong93@icloud.com}},
Kai Zhu\orcidlink{0000-0002-2583-2669}$^{2,3,4}$,
Michele Cappellari\orcidlink{0000-0002-1283-8420}$^{5}$,
Ran Li\orcidlink{0000-0003-3899-0612}$^{2,3,4}$,
Shude Mao\orcidlink{0000-0001-8317-2788}$^{1}$,
Dandan Xu$^{1}$
\\
\\
$^{1}$Department of Astronomy, Tsinghua University, Beijing 100084, China\\
$^{2}$National Astronomical Observatories, Chinese Academy of Sciences, 20A Datun Road, Chaoyang District, Beijing 100101, China\\
$^{3}$Institute for Frontiers in Astronomy and Astrophysics, Beijing Normal University, Beijing 102206, China\\
$^{4}$School of Astronomy and Space Science, University of Chinese Academy of Sciences, Beijing 100049, China\\
$^{5}$Sub-department of Astrophysics, Department of Physics, University of Oxford, Denys Wilkinson Building, Keble Road, Oxford, OX1 3RH, UK
}
\date{Accepted 22 April 2024}
\pubyear{2024}

\begin{document}
\label{firstpage}
\pagerange{\pageref{firstpage}--\pageref{lastpage}}
\maketitle
\begin{abstract}
Using the final MaNGA sample of 10K galaxies, we investigate the dark matter fraction $f_{\rm DM}$ within one half-light radius $R_{\rm e}$ for about 6K galaxies with good kinematics spanning a wide range of morphologies and stellar velocity dispersion. We employ two techniques to estimate $f_{\rm DM}$: (i) Jeans Anisotropic Modelling (JAM), which performs dark matter decomposition based on stellar kinematics and (ii) comparing the total dynamical mass-to-light ratios $(M/L)_{\rm JAM}$ and $(M_{\ast}/L)_{\rm SPS}$ from Stellar Population Synthesis (SPS). We find that both methods consistently show a significant trend of increasing $f_{\rm DM}$ with decreasing $\sigma_{\rm e}$ and low $f_{\rm DM}$ at larger $\sigma_{\rm e}$. For 235 early-type galaxies with the best models, we explore the variation of stellar initial mass function (IMF) by comparing the stellar mass-to-light ratios from JAM and SPS. We confirm that the stellar mass excess factor $\alpha_{\rm IMF}$ increases with $\sigma_{\rm e}$, consistent with previous studies that reported a transition from Chabrier-like to Salpeter IMF among galaxies. We show that the $\alpha_{\rm IMF}$ trend cannot be driven by $M_{\ast}/L$ or IMF gradients as it persists when allowing for radial gradients in our model. We find no evidence for the total $M/L$ increasing toward the centre. We detect weak positive correlations between $\alpha_{\rm IMF}$ and age, but no correlations with metallicity. We stack galaxy spectra according to their $\alpha_{\rm IMF}$ to search for differences in IMF-sensitive spectral features (e.g. the $\rm Na_{\rm I}$ doublet). We only find marginal evidence for such differences, which casts doubt on the validity of one or both methods to measure the IMF.
\end{abstract}


\section{Introduction}
\label{sec:introduction}
Stellar mass of galaxies is a key property which encodes the information of galaxy formation and evolution. By comparing the stars initially formed and those present-day observed, we are able to constrain the detailed evolution tracks of the galaxies. Thus, the stellar initial mass function (IMF), which describes the number distribution of newly formed stars of different masses, is a crucial factor to reveal the stellar mass assembly history of the galaxies. The first attempt to describe the IMF is \citet{Salpeter1955}, which proposed a single power law of the number distribution of newly-formed stars of different masses at birth, i.e. $\phi \propto m^{-2.35}$. Later on, \citet{Kroupa(2001)} and \citet{Chabrier(2003)} reported a flatter slope at low mass end (i.e. lower than $0.5\,\mathrm{M_{\odot}}$) of the IMF for the Milky Way (MW). 

To study the IMF of the MW, one can make use of direct stellar counting (e.g. \citealt{Kroupa_et_al.(1993),Phelps_et_al.(1993),Reid_et_al.(1997),Carraro_et_al.(2005),Bonatto_et_al.(2007)}). \citet{Kroupa_et_al.(1993)} analyzed the luminosity functions of a sample of stars within $5.2\,\rm pc$ of the sun and a magnitude-limited sample of more distant stars and concluded that the two samples can be described by a single broken power-law IMF, which becomes the basis of the \citet{Kroupa(2001)} IMF. \citet{Phelps_et_al.(1993)} summarized the studies on MW IMF with open clusters and found that most open clusters agree with the Salpeter IMF at the mass range of $1-8\,\mathrm{M_{\odot}}$. At the sub-solar mass range, \citet{Moraux_et_al.(2003)} studied the IMF slope down to $0.03\,\mathrm{M_{\odot}}$ with the Pleaides and found that a log-normal form can well describe the IMF over a broad mass range ($0.03-10\,\mathrm{M_{\odot}}$). Combining a large number of literature, \citet{Bastian2010} claimed in their review that there was little compelling evidence for variations in the IMF across different galaxies or within our own about ten years ago. More recently, \citet{Li_et_al.(2023)} pointed out that the IMF inside the MW varies with metallicity and cosmic time, by counting 93000 spectroscopically observed M-dwarf stars in the solar neighbourhood. So far, whether the IMF of MW varies is still an important question to address.

For external galaxies, however, one can hardly resolve single stars, and thus, cannot estimate the IMF from direct stellar counts as done in the MW. There are two main methods to study the IMF of external galaxies. One is the spectroscopic approach, which makes use of the IMF-related spectral features or the full spectrum to constrain the IMF shape of galaxies. The other is to compare the stellar mass (or stellar mass-to-light ratio) derived in a luminosity-independent (and hence, IMF-independent) way and that derived with stellar population synthesis (SPS), assuming a standard IMF. The luminosity-independent stellar mass (stellar mass-to-light ratio) can be obtained either by stellar dynamical modelling or by gravitational lensing. 

Earlier studies used the spectral features to point out that either the IMF or the chemical abundance may be different among galaxies, but they could not distinguish between the two effects \citep[e.g.][]{Faber1980,Carter1986,Cenarro2003}. Similarly, initial studies using galaxy dynamics \citep{Cappellari2006} or gravitational lensing \citep{Treu_et_al.(2010)} noted that the central $M/L$ variations among galaxies could not all be attributed to variations of the stellar population $M/L$ alone, but could not distinguish between the effects of dark matter or the IMF. Two independent studies first made a convincing case for a heavier-than-MW IMF using either strong lensing \citep{Auger2010} or spectral features \citep{van_Dokkum_et_al.(2010)}. The latter showed the IMF, and not chemical abundance, was responsible for the observed spectra variations. They reported that the gravity-sensitive spectral features (i.e. the $\mathrm{Na_I}$ doublet and the Wing–Ford molecular $\mathrm{FeH}$ band) can be detected in the spectra of the most massive elliptical galaxies, indicating the high richness of dwarf stars and hence pointing out the IMF more bottom-heavy than Salpeter in those massive elliptical galaxies (i.e. steeper slope at $m<0.3\,\mathrm{M_{\odot}}$). 

Using the galaxy-dynamics approach, \citet{Cappellari_et_al.(2012)} made use of the Jeans Anisotropic Modelling (JAM; \citealt{Cappellari2008,Cappellari2020}) to get the luminosity-independent stellar mass-to-light ratio of 260 early-type galaxies from the ATLAS$^{\rm 3D}$ survey \citep{Cappellari_et_al.(2011)} and concluded that the previously reported Salpeter-like IMF in massive galaxies was just the extreme of a systematic trend with the stellar IMF being Chabrier-like for low-mass ETGs and becoming heavier with increasing stellar mass-to-light ratio or velocity dispersion of galaxies. More details of the effort were given in \citet{Cappellari_et_al.(2013b)} who also showed that the IMF correlated better with velocity dispersion than with the mass or size of galaxies.

Following the initial reports, several subsequent studies corroborated the findings, generally supporting the original outcomes. \citet{Spiniello2012,Spiniello2015}, \citet{Conroy_et_al.(2012)} and \citet{van_Dokkum_et_al.(2012)} reported a trend between IMF and stellar velocity dispersion using the spectral-features approach. \citet{La_Barbera_et_al.(2013)} used spectral features to study the IMF variation against velocity dispersion of galaxies with 24781 early-type galaxies from the SPIDER sample \citep{La_Barbera_et_al.(2010)} and found that IMF varies from a Kroupa/Chabrier case towards a more bottom-heavy IMF with increasing central velocity dispersion. \citet{Zhou_et_al.(2019)} adopted a full spectrum analysis on the IMF variation among galaxies from the MaNGA survey \citep{Bundy_et_al.(2015)} and found the similar IMF trend with velocity dispersion. Other IMF-relevant studies with the spectroscopic approach include \citet{Ferreras_et_al.(2013),Martin-Navarro_et_al.(2015a),Martin-Navarro_et_al.(2015b),van_Dokkum_et_al.(2017),Parikh_et_al.(2018),Gu_et_al.(2022)}. \citet{Dutton_et_al.(2013)} studied the IMF in the bulges of massive spiral galaxies by comparing the stellar mass-to-light ratio from strong lensing and gas kinematics, and that from stellar population synthesis. Other studies on IMF variation with such method include \citet{Thomas_et_al.(2011),Li_et_al.(2017), Shetty_et_al.(2020),Loubser_et_al.(2021)}.
The readers are referred to \citet[sec.~4.2.2]{Cappellari(2016)} for a summary of the earlier IMF results and to \citet{Smith(2020)} for a more extended review of the IMF variation among external galaxies.

Although both the spectroscopic method and the dynamical/lensing method agree with each other on the IMF variation with velocity dispersion of galaxies, there are conflicts between the two methods. \citet{Conroy_et_al.(2012)} found that the IMF becomes more bottom-heavy with increasing $\rm [Mg/Fe]$ (similar results also see \citealt{Gu_et_al.(2022)}), although \citet{La_Barbera_et_al.(2015)} claimed that this correlation is driven by the IMF-dispersion relation and no extra correlation between IMF and $\rm [Mg/Fe]$ has been seen with the similar method. \citet{Zhou_et_al.(2019)}, with a full spectrum analysis, also found similar trend of IMF shape against total metallicity (i.e. $[Z/H]$) of galaxies. However, with a dynamical modelling-based method, \citet{Thomas_et_al.(2011)} and \citet{McDermid_et_al.(2014)} found no obvious relation between IMF and metallicity of galaxies. \citet{Smith(2014)} made a detailed comparison of the IMF shape constrained by dynamical modelling-based method and the spectroscopic method on the same galaxy sample. They found that although both methods recover the IMF-velocity dispersion trend, there is no correlation between the stellar mass excess factor inferred from the two approaches on a galaxy-by-galaxy basis. They commented that one (or both) of the methods may have not accounted fully for the main confounding factors of IMF variation and suggested that a comparison between the IMF constrained with the two methods in the same aperture would be necessary to further understand the difference.

With a series of papers of the MaNGA DynPop (Dynamics and stellar Population) project, we aim to not only provide catalogues containing the dynamical and stellar population properties for the final sample of the MaNGA survey ($\sim 10000$ nearby galaxies; \citealt{MaNGA_dr17}) as legacy products, but also to study the formation and evolution of galaxies with the combination of the dynamical and stellar population properties. In Paper I \citep{Zhu_et_al.(2023a)}, we performed the Jeans Anisotropic Modelling (JAM; \citealt{Cappellari2008,Cappellari2020}) on the full sample of MaNGA survey and obtained their quality-assessed dynamical properties. In Paper II \citep{Lu_et_al.(2023a)}, we performed stellar population synthesis (SPS) on the final MaNGA sample and obtained their global and spatially resolved stellar population properties, as well as their star-formation histories. 

This paper is the 5th paper of our MaNGA DynPop series. The goal of this paper is to understand the stellar initial mass function variation among galaxies. However, one can hardly study the IMF without understanding the dark matter content. Thus, in this work, we will combine the studies of the relation between the dark matter fraction and the stellar velocity dispersion, as well as the stellar initial mass function variation among galaxies from the currently largest integral field unit (IFU) spectroscope survey, MaNGA \citep{MaNGA_dr17}. We will employ the dynamical modelling-based method to study the IMF variation by combining the dynamical modelling-based stellar mass-to-light ratio from \citetalias{Zhu_et_al.(2023a)} and the stellar population synthesis-based stellar mass-to-light ratio from \citetalias{Lu_et_al.(2023a)}. Further, we will investigate whether the IMF variation signal inferred with the dynamical method can be confirmed with a spectral analysis of IMF. The readers are also referred to Paper III \citep{Zhu_et_al.(2024)} for a study of multiple dynamical scaling relations of these galaxies, Paper IV \citep{Wang_et_al.(2024)} for a study of the density profiles of galaxy groups and clusters, combining the stellar dynamical modelling and weak lensing, and Paper VI \citep{Li_et_al.(2024)} for a detailed comparison of total density slopes of galaxies between MaNGA and cosmological simulations.

This paper is organized as follows: in \autoref{sec:data_method}, we introduce the MaNGA project (\autoref{sec:manga}), the structural and dynamical properties (\autoref{sec:jam}), the stellar population properties (\autoref{sec:sps}) used in this paper, as well as the sample selection criteria which generate our final sample in this work (\autoref{sec:sample}). Results on dark-stellar decomposition, initial mass function variation, and IMF-sensitive spectral feature analyses are shown in \autoref{sec:dm-stellar}, \autoref{sec:imf}, and \autoref{sec:spectra}, respectively. Finally, we present our discussions and conclusions in \autoref{sec:discussion} and \autoref{sec:conclusion}, respectively.

\section{Data and Method}
\label{sec:data_method}
\subsection{MaNGA project}
\label{sec:manga}
MaNGA (Mapping Nearby Galaxies at Apache Point Observatory; \citealt{Bundy_et_al.(2015)}) currently stands as the largest integral field units (IFU) spectroscope survey in the world. MaNGA targets over 10,000 nearby galaxies within the redshift range of $0.01<z<0.15$ \citep{Yan_et_al.(2016b),Wake_et_al.(2017)}, with the wavelength range of the spectra spanning from $3600\,\Angstrom$ to $10300\,\Angstrom$ and a spectral resolution of about $R\sim 2000$ (\citealt{Drory_et_al.(2015)}; see \citealt{Yan_et_al.(2016a)} for more details on MaNGA spectrophotometry calibration). Data products of the spectra are generated from the calibration and reduction on the raw data using the Data Reduction Pipeline (DRP; \citealt{Law_et_al.(2016)}). 67\% of the galaxies are observed out to $1.5R_{\rm e}$ (where $R_{\rm e}$ is the effective radius of galaxies) and 33\% are observed out to $2.5R_{\rm e}$, conducting the "Primary+" and "Secondary" samples, respectively \citep{Bundy_et_al.(2015)}. Readers are referred to the following papers for more details on the MaNGA instrumentation \citep{Drory_et_al.(2015)}, observing strategy \citep{Law_et_al.(2015)}, spectrophotometric calibration \citep{Smee_et_al.(2013),Yan_et_al.(2016a)}, and survey execution and initial data quality \citep{Yan_et_al.(2016b)}. The MaNGA project is now finished and the complete data have been released (\citealt{MaNGA_dr17}).

\subsection{The axisymmetric Jeans Anisotropic Modelling}
\label{sec:jam}
We obtain the structural and dynamical properties of the full MaNGA sample from the first paper of our MaNGA DynPop series (Paper I; \citealt{Zhu_et_al.(2023a)}). In \citetalias{Zhu_et_al.(2023a)}, the axisymmetric Jeans Anisotropic Modelling (JAM; \citealt{Cappellari2008,Cappellari2020}) is applied to the complete sample of MaNGA to extract their quality-assessed structural and dynamical properties, using the {\sc python} version of JAM, {\sc JamPy}\footnote{Version 6.3, available from \url{https://pypi.org/project/jampy/}}. 

\subsubsection{JAM mass models}
\label{sec:jam_models}
\citetalias{Zhu_et_al.(2023a)} provides the properties from JAM with different assumptions on the shape of dark halos and velocity dispersion ellipsoids. In this work, we adopt the results with the cylindrically-aligned velocity dispersion ellipsoid \citep{Cappellari2008} and compare the results under different mass model assumptions. The mass models adopted in this work include:
\begin{enumerate}
\item The mass-follows-light model (MFL model hereafter; also know as the self-consistent model, e.g. \citealt{Cappellari_et_al.(2013a),Shetty_et_al.(2020)}): The total mass distribution is assumed to follow the SDSS $r$-band \citep{Stoughton_et_al.(2002)} luminosity distribution of galaxies and no stellar-dark matter (DM) decomposition is applied.

\item JAM model with an NFW dark halo (NFW model hereafter): In this model, the total mass distribution consists with a oblate stellar component (following the SDSS $r$-band luminosity distribution by assuming a constant stellar mass-to-light ratio) and a spherical NFW dark halo \citep{Navarro_et_al.(1997)}.

\item JAM model with a gNFW dark halo (gNFW model hereafter): This model is similar to the NFW model, but the dark halo is assumed to be a spherical gNFW model \citep{Wyithe2001}, which allows for the variation of the inner density slope $\gamma$ of the dark halo (in NFW dark halo, it is set to be $\gamma=-1$).

\end{enumerate}

In all JAM models, the supermassive black hole (SMBH) mass of galaxies is estimated with the $M_{\rm BH}-\sigma_{\rm c}$ relation from \citet{McConnell_et_al.(2011)}, where $M_{\rm BH}$ is the mass of the SMBH and $\sigma_{\rm c}$ is the stellar velocity dispersion within the aperture with radius being the full width at half maximum (FWHM) of the MaNGA point spread function (PSF).

\subsubsection{Structural and dynamical properties}
Below, we briefly introduce the structural and dynamical properties of galaxies used in this work, all of which are taken from the catalogue of \citetalias{Zhu_et_al.(2023a)}.
\begin{enumerate}
    \item Half-light radii: $R_{\rm e}^{\rm maj}$ is the semi-major axis of the elliptical half-light isophotes of the galaxies. $R_{\rm e}$ satisfies $\pi R_{\rm e}^2  = A$, where $A$ is the area of the elliptical half-light isophote. The elliptical half-light isophote (i.e. the galaxy centre, the global ellipticity, and the sizes) is obtained with the {\sc python} software, {\sc MgeFit}\footnote{Version 5.0.14, available from \url{https://pypi.org/project/mgefit/}} by \citet{Cappellari(2002)}. The sizes are scaled by a factor of 1.35 to approximately match the popular galaxy sizes obtained from extrapolated photometry (see fig.~7 of \citealt{Cappellari_et_al.(2013b)}).
    
    \item $\sigma_{\rm e}$ is the luminosity-weighted velocity dispersion calculated within the elliptical half-light isophote. Specifically, it is calculated as:
    \begin{equation}
    \label{eq:sigma_e}
    \sigma_{\rm e} = \sqrt{\frac{\sum_{k}F_{k}(V_{k}^2+\sigma_{k}^2)}{\sum_{k}F_{k}}},
    \end{equation}
    where $V_{k}$ and $\sigma_{k}$ are the mean line-of-sight velocity and dispersion in the $k$-th spaxel, respectively, and $F_{k}$ is the corresponding flux. The summation goes over all the spaxels within the elliptical half-light isophote. Velocities and dispersion here are obtained by the MaNGA Data Analysis Pipeline (DAP; \citealt{Belfiore_et_al.(2019),Westfall_et_al.(2019)}).

    \item $(M/L)_{\rm JAM}$ is the total mass-to-light ratio (including stellar and dark matter) of the galaxies. 
    
    \item $(M_{\ast}/L)_{\rm JAM}$ is the stellar mass-to-light ratio of the galaxies.
    

    \item $f_{\rm DM}$ is the dark matter fraction within a sphere of radius $R_{\rm e}$.
\end{enumerate}

In this work, we take the total mass-to-light ratio from the MFL model and the other JAM-related properties (i.e. JAM-based stellar mass-to-light ratio and JAM-based dark matter fraction) from the NFW model. The NFW model is set as the default model of this work and the comparisons between JAM models are made in \autoref{sec:imf2sigma_JAMmodels}. The readers are referred to sec.~3.3 of \citetalias{Zhu_et_al.(2023a)} for more details on mass model design of JAM.

\subsection{Stellar population properties of galaxies}
\label{sec:sps}
The stellar population properties used in this work is derived from the catalogue\footnote{Available from \url{https://github.com/manga-dynpop}} in the second paper of our MaNGA DynPop series (Paper II; \citealt{Lu_et_al.(2023a)}). \citetalias{Lu_et_al.(2023a)} performs the Penalized Pixel-Fitting method ({\sc ppxf}\footnote{Version 8.2, available from \url{https://pypi.org/project/ppxf/}.}; \citealt{Cappellari_et_al.(2004),Cappellari2017,Cappellari(2023)}) on the spectra produced by MaNGA DRP. As pointed out by \citet{Woo_et_al.(2024)}, {\sc ppxf} is 3-4 times faster than other SPS codes, including {\sc firefly} \citep{Wilkinson_et_al.(2017)}, {\sc pypipe3d} \citep{Sanchez_et_al.(2016a),Sanchez_et_al.(2016b),Lacerda_et_al.(2022)}, and {\sc starlight} \citep{Cid_Fernandes_et_al.(2005)}, and is the best in recovering stellar population properties among the four softwares. The simple stellar population (SSP) model used in \citetalias{Lu_et_al.(2023a)} is generated with the {\sc fsps}\footnote{Available from \url{https://github.com/cconroy20/fsps}.} software \citep{Conroy_et_al.(2009),Conroy_et_al.(2010)}, combining the \citet{Salpeter1955} IMF and the MIST isochrones \citep{Choi_et_al.(2016)}. Specifically, the relevant stellar population properties in this work include:
\begin{enumerate}
\item $\lg\,\mathrm{Age}$ (keyword: \texttt{LW\_Age\_Re} from the catalogue of \citetalias{Lu_et_al.(2023a)}): $\lg\,\mathrm{Age}$ is the global luminosity-weighted age of the galaxies, which is derived from the {\sc ppxf} fitting on the stacked spectra within the elliptical half-light isophotes. With the best-fitted weights from {\sc ppxf}, the luminosity-weighted age is defined as:
\begin{equation}
    \label{eq:lwsp}
    x = \frac{\sum_{k} w_k L_k x_k}{\sum_k w_k L_k},
\end{equation}
where $x$ here is $\lg\, \mathrm{Age}$; $w_k$ and $L_k$ are the best-fitted weight and the corresponding SDSS $r$-band luminosity of the $k-$th template, respectively.

\item $[Z/H]$ (keyword: \texttt{LW\_Metal\_Re}): $[Z/H]$ is the global luminosity-weighted metallicity of the galaxies, derived with the same way as the global age.

\item $(M_{\ast}/L)_{\rm SPS}$ (keyword: \texttt{ML\_int\_Re}): $(M_{\ast}/L)_{\rm SPS}$ used here is stellar mass-to-light ratio of galaxies, fitted from the stacked spectra within the elliptical half-light isophotes. It is defined as:
\begin{equation}
    \label{eq:ml}
    (M_{\ast}/L)_{\rm SPS} = \frac{\sum_k  w_k M_{\ast,k}}{\sum_k w_k L_k},
\end{equation}
where $M_{\ast,k}$ is the stellar mass (including the mass of living stars and stellar remnants, but excluding the gas lost during stellar evolution) and $L_k$ is the corresponding luminosity of $k-$th templates. This quantity is {\em intrinsic} to the stellar population, being a unique function of the age and metallicity distribution of the stellar population. For this reason it is independent of attenuation effects, which are accounted for during the \textsc{ppxf} fit. The summation goes over all the templates used in the fitting.

\item $A_V$ (keyword: \texttt{Av\_Re}): $A_V$ is the average dust attenuation effect at $\lambda = 5500\,\Angstrom$ ($V$-band) within the elliptical half-light isophote.

\item $f_{L_r}$ (keyword: \texttt{Fred\_tot\_Re}): $f_{L_r}$ is the $r-$band luminosity ratio between the {\it observed} spectra and the {\it intrinsic} spectra, both of which are stacked within the elliptical half-light isophotes (see fig.~2 of \citetalias{Lu_et_al.(2023a)} for details).

\item $\gamma_{M_{\ast}/L}$ (keyword: \texttt{ML\_int\_Slope}): $\gamma_{M_{\ast}/L}$ is the gradient of {\em intrinsic} $r-$band stellar mass-to-light ratio within the elliptical half-light isophote.

\end{enumerate}

We note here that although the mass-to-light ratios taken from \citetalias{Zhu_et_al.(2023a)} and \citetalias{Lu_et_al.(2023a)} are measured in the same aperture, they are not comparable without further calibration. The SPS-based stellar mass-to-light ratio (derived from \citetalias{Lu_et_al.(2023a)}) is defined to be the ``intrinsic'' values (i.e. dust attenuation effect independent; see \citetalias{Lu_et_al.(2023a)} for more details), while the JAM-based mass-to-light ratio (derived from \citetalias{Zhu_et_al.(2023a)}) is estimated without correcting for the dust attenuation effect. Thus, before comparing the mass-to-light ratios from different methods and further constraining the initial mass function, we correct the JAM-based (total or stellar) mass-to-light ratio using the dust attenuation factor estimated from \citetalias{Lu_et_al.(2023a)}, with:
\begin{equation}
(M/L)_{\rm JAM, corr.} = (M/L)_{\rm JAM, uncorr.}\times f_{L_r},
\end{equation}
where $(M/L)_{\rm JAM, uncorr.}$ is the uncorrected JAM-based mass-to-light ratio (either stellar or total mass-to-light ratios), directly taken from the catalogue of \citetalias{Zhu_et_al.(2023a)}. $f_{L_r}$ is the estimate of dust attenuation effect given by {\sc ppxf} fitting (see \citetalias{Lu_et_al.(2023a)} for details). For simplicity, in following sections, the symbols, $(M_{\ast}/L)_{\rm JAM}$ and $(M/L)_{\rm JAM}$ are used only for the ``corrected'' JAM-based stellar and total mass-to-light ratios, respectively. We emphasize here that such a reddening correction is only an approximation. The reddening effect is small, which only reduces the JAM-based mass-to-light ratios (and hence the stellar mass excess factor; see \autoref{sec:imf2sigma} for definition) by $\rm 0.13\, dex$ (median value) for the $\rm Qual\geqslant 1$ sample used in \autoref{sec:dm-stellar} and $0.08\,\rm dex$ for the high-quality early-type galaxies used in \autoref{sec:imf} and \autoref{sec:spectra} (see \autoref{sec:sample} for sample selection criteria). Thus, the choice of reddening correction method will not qualitatively change the results in this work.

\subsection{Sample selection}
\label{sec:sample}
Combining the two catalogues from \citet{Zhu_et_al.(2023a)} (Paper I) and \citet{Lu_et_al.(2023a)} (Paper II), we get 9878 unique galaxies with available dynamical and stellar population properties. As stated in \citetalias{Zhu_et_al.(2023a)}, galaxies are assigned with a quality flag, which indicates the reliability of the dynamical quantities (see table~2 of \citetalias{Zhu_et_al.(2023a)} for details). In \autoref{sec:dm-stellar} of this work, we only take galaxies with $\mathrm{Qual}\geqslant 1$ (total mass estimates can be trusted) into account, which results in 5952 galaxies with different morphologies, in order to balance the sample size and the parameter reliability. In IMF-related analyses (i.e. \autoref{sec:imf} and \autoref{sec:spectra}), however, a stricter criteria should be applied. Firstly, we only take galaxies with $\mathrm{Qual}= 3$ (the highest quality) into account, which have the most reliable estimate of JAM-based stellar mass-to-light ratio of galaxies, resulting in 1129 galaxies with different morphologies. Secondly, we select the galaxies that satisfy:
\begin{equation}
\Delta \lg\,(M_{\ast}/L)_{\rm JAM} = \left|\lg (M_{\ast}/L)_{\rm JAM}^{\rm cyl}-\lg (M_{\ast}/L)_{\rm JAM}^{\rm sph}\right|<0.05,
\end{equation}
where $(M_{\ast}/L)_{\rm JAM}^{\rm cyl}$ and $(M_{\ast}/L)_{\rm JAM}^{\rm sph}$ are the stellar mass-to-light ratios from JAM with cylindrically- \citep{Cappellari2008} and spherically-aligned \citep{Cappellari2020} velocity dispersion ellipsoids, respectively, further resulting in 909 galaxies (we name these galaxies as the ``golden sample'') out of 1142 $\mathrm{Qual}=3$ galaxies.

Measuring the IMF in external spiral galaxies is more challenging than in ETGs because the spectra of spirals are dominated by bright stars in their young stellar population. This makes it difficult to quantify the contribution of faint low-mass stars from their spectral features. Additionally, spiral galaxies contain large amounts of dust and multiple star formation events, which make the stellar population $M_*/L$ more uncertain. As a result, both spectral-features and dynamics/lensing methods are less reliable for measuring the IMF in spiral galaxies. For these reasons, most previous studies on IMF variation focus only on early-type or passive galaxies for both the spectral-features method (e.g. \citealt{van_Dokkum_et_al.(2010),Spiniello2012}) and the dynamical \citep[e.g.][]{Cappellari_et_al.(2012),Shetty_et_al.(2020)} or lensing \citep[e,g,][]{Treu_et_al.(2010),Auger2010} methods. An exception is the work by \citet{Li_et_al.(2017)}, which tried to correct for dust and gas in MaNGA spiral galaxies with some empirical relations. 

In this work, we also focus on early-type galaxies (ETGs) in all our IMF-related analyses, except \autoref{sec:comparison_sample}, where we show the comparison of IMF$-\sigma_{\rm e}$ relation under different sample selection criteria. To classify early- and late-type galaxies, we employ the morphology catalogue from \citet{Dominguez_Sanchez_et_al.(2022)}, which applied a deep learning method to obtain the classification of morphologies of the full sample of MaNGA. \citet{Dominguez_Sanchez_et_al.(2022)} provided multiple classification criteria to select early-type galaxies. Here, we simply take the galaxies with $\mathrm{T-Type}\leqslant 0$ as early-type galaxies, resulting in 235 out of 909 galaxies in the golden sample. We confirm that our results would keep nearly unchanged when different morphology classification criteria are applied. In the following sections, we name this sample as the ``morphology-based sample''.

\section{Stellar mass-to-light ratio and dark matter fraction}
\label{sec:dm-stellar}

To study the IMF one first needs to understand the variations of the dark matter fraction, which can give a similar effect as the IMF when the total and stellar mass distribution have similar profiles. We already presented trends of the dark matter fraction $f_{\rm DM}(<R_{\rm e})$ in \citetalias{Zhu_et_al.(2024)} as a function of stellar mass $M_{\ast}$. We also showed in fig.~16 therein that $f_{\rm DM}(<R_{\rm e})$ best correlates with $\sigma_{\rm e}$ rather than $M_{\ast}$. Here we study the trends as a function of $\sigma_{\rm e}$ directly. Moreover, in addition to using the dark matter fraction inferred with JAM, we also make use of an alternative approach, based on comparing stellar population and dynamics, to strengthen our results.

\subsection{Mass-to-light ratio vs. velocity dispersion}

\begin{figure*}
	\centering
	\includegraphics[width=2\columnwidth]{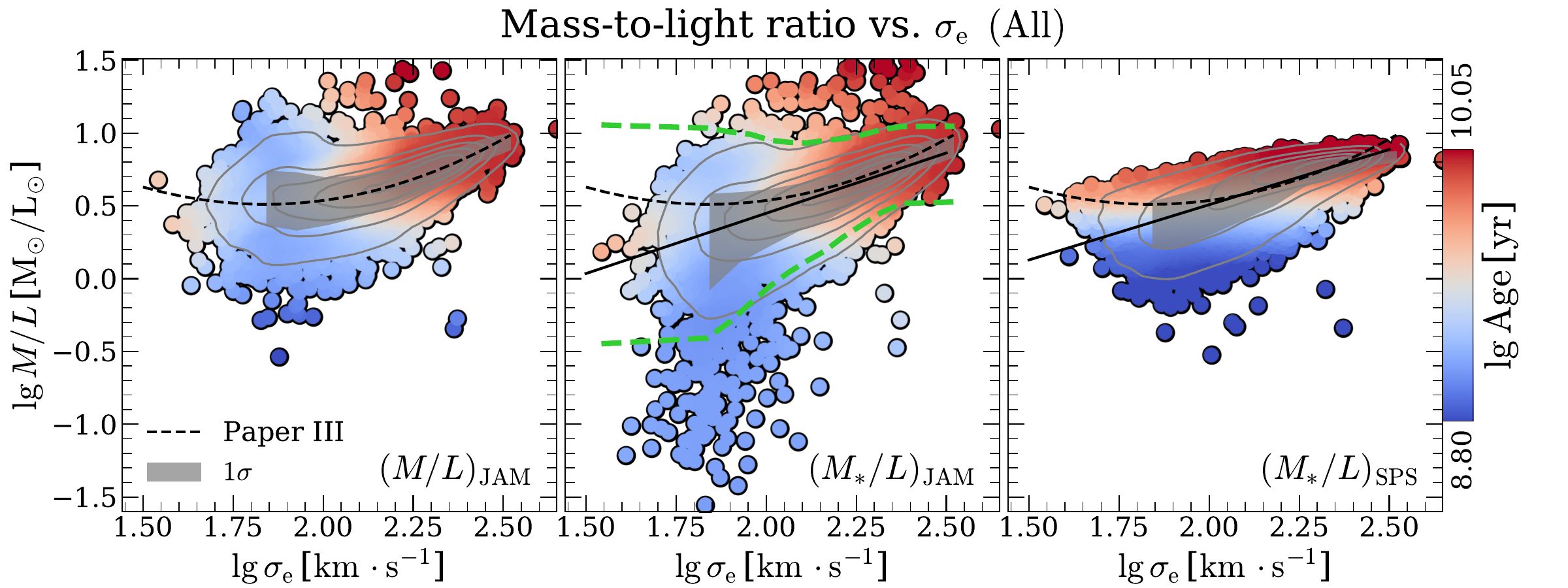}
    \includegraphics[width=2\columnwidth]{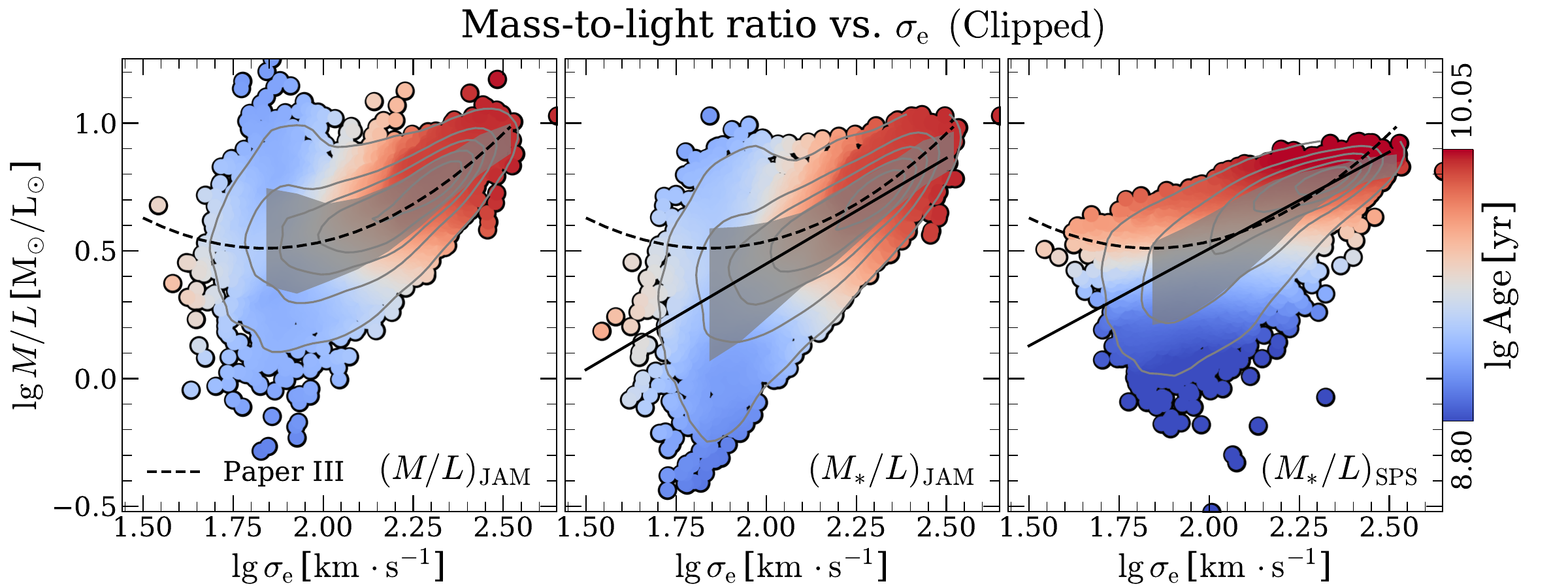}
	\caption{Top: the correlation between mass-to-light ratio and velocity dispersion of galaxies. Here we included both spirals and ETGs models with $\rm Qual>0$. From left to right, we show the trend of (1) total (including stars and dark matter) mass-to-light ratio obtained with an MFL JAM model (see \autoref{sec:jam_models} for details of the models); (2) the stellar mass-to-light ratio obtained with an NFW JAM model; (3) the stellar mass-to-light ratio within the elliptical half-light isophote obtained with SPS. In each panel, the colours indicate the global luminosity-weighted age within the elliptical half-light isophote (see \autoref{sec:sps} for details). The black dashed curve indicates the best-fitted parabolic relation between dynamical mass-to-light ratio and $\lg\,\sigma_{\rm e}$ from \citetalias{Zhu_et_al.(2024)}. The solid black line indicates the best-fitted linear correlation between the stellar mass-to-light ratio and the velocity dispersion. The grey shaded regions indicate the range from 16 to 84 percentiles ($\pm 1\sigma$). The green dashed curves in the middle panel indicate the $3\sigma$ region, out of which the galaxies are clipped. A kernel density estimation of the galaxy number density is indicated by the grey contours. Bottom: the same as the top sub-figure but the samples shown are clipped based on the $\lg\,(M_{\ast}/L)_{\rm JAM}-\lg\,\sigma_{\rm e}$ relation to exclude the outliers out of the $3\sigma$ region using the {\sc ndimage.percentile\_filter} routine of the {\sc scipy} software.}
	\label{fig:ml2sigma}
\end{figure*}

In the top sub-figure of \autoref{fig:ml2sigma}, we present the distributions of three different mass-to-light ratios as a function of $\sigma_{\rm e}$ from left to right for all the MaNGA galaxies with $\rm Qual\geqslant1$ (5952 galaxies\footnote{There are 6065 galaxies with $\rm Qual\geqslant1$ according to \citet[][table 2]{Zhu_et_al.(2023a)}, among which 5952 have available SPS-based stellar mass-to-light ratio estimates.}). The first ratio is the total mass-to-light ratio $(M/L)_{\rm JAM}$ inside a sphere of radius $R_{\rm e}$, which includes both stellar and dark matter and is obtained from JAM with an MFL model (see \autoref{sec:jam_models} for details of JAM model design). The second ratio is the stellar mass-to-light ratio $(M_*/L)_{\rm JAM}$, which is obtained from JAM with an NFW model. The third ratio is the average SPS-based stellar mass-to-light ratio $(M_*/L)_{\rm SPS}$ within the elliptical half-light isophote.

The left panel of the top sub-figure of \autoref{fig:ml2sigma} is the same as the top panel of fig.~5 in \citetalias{Zhu_et_al.(2024)}, where a clear parabolic relation between $\lg\,(M/L)_{\rm JAM}$ and $\lg\,\sigma_{\rm e}$ is seen. The total mass-to-light ratio first slightly decreases with $\sigma_{\rm e}$ until $\lg\,(\sigma_{\rm e}/\mathrm{km\,s^{-1}})\sim 1.89$ (from the best-fitted correlation in \citetalias{Zhu_et_al.(2024)}), and then increases with $\sigma_{\rm e}$. The JAM-based stellar mass-to-light ratio, $\lg\,(M_{\ast}/L)_{\rm JAM}$, however, does not show the same parabolic relation with $\lg\,\sigma_{\rm e}$ as the total mass-to-light ratio $\lg\,(M/L)_{\rm JAM}$, but shows roughly linear correlation with $\lg\sigma_{\rm e}$ (see the contours of its distribution and the figure for clipped sample in the bottom sub-figure). At the low $\sigma_{\rm e}$ end, $(M/L)_{\rm JAM}$ is significantly higher than $(M_{\ast}/L)_{\rm JAM}$, compared to larger velocity dispersions. This indicates that the dark matter in the inner region of galaxies (within about $1R_{\rm e}$) is more important for low $\sigma_{\rm e}$ galaxies. With increasing $\sigma_{\rm e}$, the difference between total mass-to-light ratio and JAM-based stellar mass-to-light ratio becomes smaller, until $\lg\,\sigma_{\rm e}\sim 2.25$, indicating the decrease of dark matter content in the inner region of the galaxies. 

Crucially, the same approximately linear trend of logarithmic quantities (i.e. power-law relation) is also seen for the SPS-based stellar mass-to-light ratio in the right panel of \autoref{fig:ml2sigma} (top). The consistent lack of a parabolic trend in both the $\lg\,(M_{\ast}/L)_{\rm SPS}-\lg\,\sigma_{\rm e}$ and $\lg\,(M_{\ast}/L)_{\rm JAM}-\lg\,\sigma_{\rm e}$ relations, combined with a clear parabolic trend in the $\lg(M/L)_{\rm JAM}-\lg\sigma_{\rm e}$ relation leaves no doubt about the reality of the trend of increasing dark matter starting below $\lg\sigma_{\rm e}\la2.1$.

It is also clear in the middle panel of \autoref{fig:ml2sigma} that there are a number of galaxies which strongly deviate from the main trend. The tail of galaxies with unrealistically-low $\lg\,(M_{\ast}/L)_{\rm JAM}\la-0.2$ must be the objects for which the dark matter is overestimated due to the degeneracy in the model parameters. The ones at the top with $\lg\,(M_{\ast}/L)_{\rm JAM}\ga1.1$ are likely face-on models, for which the stellar mass-to-light ratios tend to be overestimated \citep{Lablanche2012}. The existence of this clear $\lg\,(M_{\ast}/L)_{\rm JAM}-\lg\,\sigma_{\rm e}$ relation allows us to detect problematic models and remove them from the analysis that follows. Thus, in the bottom sub-figure of \autoref{fig:ml2sigma}, we show the same distribution as the top sub-figure but for the sample for which we have excluded the outliers out of $3\sigma$ in order to show a cleaner trend. The clipping is performed on the $\lg\,\sigma_{\rm e}-\lg\,(M_{\ast}/L)_{\rm JAM}$ relation using the {\sc ndimage.percentile\_filter} routine of the {\sc scipy} software \citep{Scipy2020} and then the same clipped galaxies in the $\lg\,\sigma_{\rm e}-\lg\,(M_{\ast}/L)_{\rm JAM}$ relation are also excluded from the other two relations. There are 349 out of 5952 $\rm Qual\geqslant1$ galaxies clipped and we can see that the main trends we see in the top sub-figure remain unchanged.

In all panels, we also show the smoothed mean luminosity-weighted age from \citetalias{Lu_et_al.(2023a)}. It was computed with the \textsc{loess\_2d} routine\footnote{We used v2.1 available from \url{https://pypi.org/project/loess/}.} of \citet{Cappellari_et_al.(2013b)}, which implements the multivariate {\sc loess} algorithm of \citet{Cleveland_and_Devlin(1988)}. It is interesting to see that both the JAM-based total mass-to-light ratio, $(M/L)_{\rm JAM}$, and the JAM-based stellar mass-to-light ratio, $(M_{\ast}/L)_{\rm JAM}$ do not show correlation with age at fixed $\sigma_{\rm e}$, while the SPS-based stellar mass-to-light ratio shows strong age dependence at fixed $\sigma_{\rm e}$. The latter is expected as we already see similar distributions of age and SPS-based stellar mass-to-light ratio on the mass-size plane (e.g. \citealt{Li_et_al.(2018)} and \citetalias{Lu_et_al.(2023a)}), which implies the tight correlation between the two parameters. \citet[][fig.~1]{Cappellari_et_al.(2013b)} pointed out that the JAM-based mass-to-light ratio also varies along the direction of $\sigma_{\rm e}$ on the mass-size plane, similar to the stellar population properties (i.e. age and SPS-based stellar mass-to-light ratio), but it is only for early-type galaxies. We confirm that late-type galaxies do not show tight correlation between age from SPS and mass-to-light ratio from JAM, causing the lack of this correlation for the whole sample as shown in \autoref{fig:ml2sigma}. This may be because that the high gas content of late-type galaxies contribute to the JAM-based mass-to-light ratio, washing out its correlation with age. 

\subsection{Dark matter vs. velocity dispersion}

\begin{figure*}
	\centering
	\includegraphics[width=1.7\columnwidth]{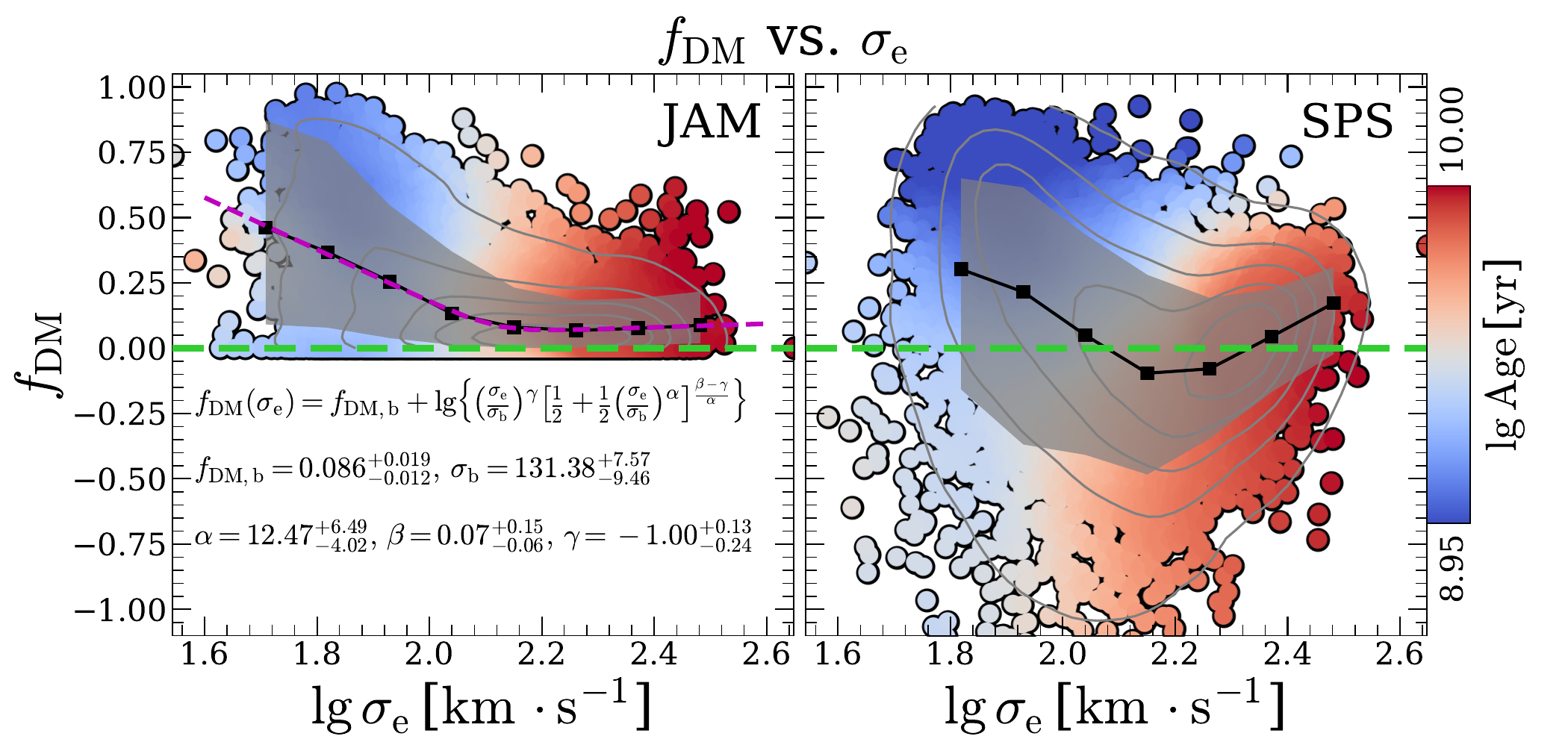}
	\caption{The correlation between dark matter fraction ($f_{\rm DM}$) and velocity dispersion of galaxies (the clipped sample from \autoref{fig:ml2sigma}). The $f_{\rm DM}$ in the left panel is from JAM with an NFW model (estimated within a sphere with radius being $R_{\rm e}$; see \autoref{sec:jam_models} for details of JAM models). The $f_{\rm DM}$ in the right panel is simply calculates as $f_{\rm DM} = 1-(M_{\ast}/L)_{\rm SPS}/(M/L)_{\rm JAM}$ (following \citealt[][eq.~22]{Cappellari_et_al.(2013a)}). In each panel, the colours indicate the global luminosity-weighted age of the galaxies (i.e. the average age within the elliptical half-light isophote; see \autoref{sec:sps} for details). The black squares and the black solid curve indicate the median trend of $f_{\rm DM}-\lg\,\sigma_{\rm e}$ relation, with the grey shaded region indicating the $1\sigma$ region. A kernel density estimation of the galaxy number density is indicated by the grey contours. The green line indicates $f_{\rm DM}=0$, below which the dark matter fraction is not applicable to JAM-based results. For the JAM-based dark matter fraction, we fit a double-power law relation to the $f_{\rm DM}-\lg\,\sigma_{\rm e}$ correlation (the magenta dashed curve) and the result is shown in the left panel.}
	\label{fig:fdm2sigma}
\end{figure*}

Having qualitatively established the need for non-luminous matter at low $\sigma_{\rm e}$, now we can try to quantify it. Thus, in \autoref{fig:fdm2sigma}, we present the correlation between dark matter fraction (either derived from JAM or SPS) and velocity dispersion. The JAM-based dark matter fraction $f_{\rm DM}^{\rm JAM}$ is calculated with the sphere with radius $R_{\rm e}$, while the SPS-based dark matter fraction is calculated with the following approximation \citep[eq.~22]{Cappellari_et_al.(2013a)}:
\begin{equation}
\label{eq:fdm_sps}
f_{\rm DM}^{\rm SPS} = 1-\frac{(M_{\ast}/L)_{\rm SPS}}{(M/L)_{\rm JAM}}.
\end{equation}
It relies on the empirical fact that $(M/L)_{\rm JAM}\approx(M/L)(<R_{\rm e })$, where $(M/L)_{\rm JAM}$ is measured from MFL models. In \citetalias{Zhu_et_al.(2023a)}, we confirmed that, especially on low-quality data where it matters most, $(M/L)_{\rm JAM}$ provides a more accurate estimate of $(M/L)(<R_{\rm e })$ than the value integrated from JAM models with dark matter.

As can be seen, the JAM-based dark matter fraction decreases with $\sigma_{\rm e}$ from $\sim 0.4$ (median value) at $\lg\,\sigma_{\rm e} \sim 1.8$ to $f_{\rm DM}<0.1$ at $\lg\,\sigma_{\rm e}\sim 2.2$ and keep nearly constant afterwards, consistent with what we see in \autoref{fig:ml2sigma}. This trend is also seen for the nearly-independent SPS-based dark matter fraction, where SPS-based dark matter fraction is also higher at low $\sigma_{\rm e}$. It reaches the lowest dark matter fraction at $\lg\sigma_{\rm e}\sim 2.2$ and then increases with increasing $\sigma_{\rm e}$. We note that in this work, a fixed Salpeter IMF is assumed when obtaining the SPS-based stellar mass-to-light ratio. It overestimates $(M_{\ast}/L)_{\rm SPS}$ at the low $\sigma_{\rm e}$ (thus underestimate the dark matter fraction) and underestimates $(M_{\ast}/L)_{\rm SPS}$ at the high $\sigma_{\rm e}$ (thus overestimate the dark matter fraction) as the IMF varies with $\sigma_{\rm e}$ (e.g. \citealt{van_Dokkum_et_al.(2010),Cappellari_et_al.(2012)} and \autoref{sec:imf} of this work). It also causes the negative SPS-based dark matter fractions at intermediate $\sigma_{\rm e}$. Although there is stellar-DM degeneracy in dynamical modelling (see \citealt{Li_et_al.(2016)} for a quantitative investigation of the degeneracy in JAM with cosmological simulation), combining the trends of dark matter fraction and $\sigma_{\rm e}$ from JAM and SPS, we are still able to confirm the decreasing trend of dark matter fraction from low to high $\sigma_{\rm e}$ on a statistical basis. Again, we see that $f_{\rm DM}^{\rm JAM}$ does not show correlation with galaxy age, while $f_{\rm DM}^{\rm SPS}$ still show age dependence even at fixed $\sigma_{\rm e}$, which may be due to the gas contribution in the JAM-based mass-to-light ratio (and hence, the dark matter fraction).

\autoref{fig:fdm2sigma} constitutes the most robust evidence so far for an increase of the dark matter fraction within $R_{\rm e}$ for decreasing velocity dispersion below $\lg\sigma_{\rm e}\la2.1$. Given that there is no evidence for the IMF to increase at low $\sigma_{\rm e}$, the increase of $f_{\rm DM}^{\rm SPS}$ provides a strong confirmation of the same trend in $f_{\rm DM}^{\rm JAM}$. The fact that $f_{\rm DM}^{\rm SPS}<f_{\rm DM}^{\rm JAM}$ must be due to the assumed Salpeter IMF which overestimates the stellar mass and consequently underestimates $f_{\rm DM}^{\rm SPS}$ at low $\sigma_{\rm e}$. We confirm that the $f_{\rm DM}^{\rm SPS}$-related results remain unchanged when $(M/L)_{\rm JAM}$ used in \autoref{eq:fdm_sps} is replaced by the total mass-to-light ratio from JAM models with dark halos (i.e. the NFW and gNFW models). Here, we fit a double-power law relation to the JAM-based $f_{\rm DM}-\lg\,\sigma_{\rm e}$ relation with the formula being:
\begin{equation}
\label{eq:fdm2sigma}
f_\mathrm{DM}(\sigma_{\rm e}) = f_\mathrm{DM,b} + \lg \left\{\left(\frac{\sigma_{\rm e}}{\sigma_{\rm b}}\right)^\gamma
\left[\frac{1}{2} + \frac{1}{2} \left(\frac{\sigma_{\rm e}}{\sigma_{\rm b}}\right)^{\alpha}\right]^{\frac{\beta - \gamma}{\alpha}}\right\},
\end{equation}
where $\sigma_{\rm b}$ is the break value of $\sigma_{\rm e}$; $f_{\rm DM,b}$ is the dark matter fraction at the break $\sigma_{\rm b}$; $\alpha$ is the sharpness; $\beta$ is the slope at large $\sigma_{\rm e}$; $\gamma$ is the slope at small $\sigma_{\rm e}$. We list the best-fitted values of the parameters in \autoref{table:table_eqfdm}, together with their measurement uncertainties. The measurement uncertainties are derived using the bootstrap method, following the steps below:
\begin{enumerate}
\item Randomly select half of the samples from the original dataset (used in \autoref{fig:fdm2sigma}).
\item Calculate the median profile for the sub-sample and fit the same double-power law relation as \autoref{eq:fdm2sigma}.
\item Repeat the above steps 500 times and calculate the 16th and 84th percentiles of the 500 best-fitted parameter sets.
\end{enumerate}

\begin{table}
  \caption{The best-fitted parameters of \autoref{eq:fdm2sigma} and their measurement uncertainties derived using the bootstrap method.} \setlength{\tabcolsep}{2.5mm}
\begin{tabular}{ccccc}
\hline
\hline
\\
 $f_{\rm DM,b}$ & $\sigma_{\rm b}$ & $\alpha$ & $\beta$ & $\gamma$\\
\\
\hline
\\
$0.086^{+0.019}_{-0.012}$ & $131.38^{+7.57}_{-9.46}$ & $12.47^{+6.49}_{-4.02}$ & $0.07^{+0.15}_{-0.06}$ & $-1.00^{+0.13}_{-0.24}$\\
\\
\hline
\end{tabular}
\vspace{2mm}
\label{table:table_eqfdm}
\end{table}

\subsection{Comparison with previous studies}
Many previous studies have been made to investigate the dark matter content in different galaxies. For example, \citet{Toloba_et_al.(2014)}, using crude virial estimates of the galaxy masses, found that the dark matter fraction of dwarf early-type (dE) galaxies in the Virgo cluster is lower than the dark matter fraction of early-type galaxies in \citet{Cappellari_et_al.(2013a)}. \citet{Tortora_et_al.(2016)} used spherical isotropic Jeans models and confirmed the results of \citet{Toloba_et_al.(2014)} and further pointed out a decreasing trend of dark matter fraction with increasing velocity dispersion (see fig.~3 therein). \citet{Eftekhari2022} used the virial estimator and found that low mass dwarf galaxies tend to have higher dynamical-to-stellar mass ratio (or equivalently, higher dark matter fraction) than high mass dwarf galaxies with a larger sample. All these studies confirms the fact that dark matter becomes more important in low mass galaxies. \citet{Eftekhari2022} interpreted it as the consequence of the stronger supernovae energy feedback and UV photoionization in lower mass halos (see also \citealt{Babul_et_al.(1992),Martin1999,Dekel_et_al.(2003)}) which lowers the star-formation efficiency therein. However, these studies are all based on the crude virial mass estimates. With accurate dynamical modelling, \citet{Cappellari_et_al.(2013a)} also found that dark matter fraction increases with decreasing stellar mass for low mass early-type ATLAS$^{\rm 3D}$ galaxies (see fig.~10 therein). \citet{Santucci2022} studied the dark matter content of SAMI galaxies \citep{Bryant_et_al.(2015)} with the Schwarzschild orbit-superposition models \citep{Schwarzschild1979} and also found the similar result (see fig.~7 therein). However, both studies focus only on the massive galaxies ($M_{\ast}\gtrsim 6\times 10^{9}\,\mathrm{M_{\odot}}$ for \citealt{Cappellari_et_al.(2013a)} and $\lg\,(M_{\ast}/\mathrm{M_{\odot}})>9.5$ for \citealt{Santucci2022}). Moreover, all previous studies studied dark matter correlations with $M_{\ast}$. But we found in \citetalias{Zhu_et_al.(2024)} that DM correlates better with stellar velocity dispersion than with stellar mass and for this reason we study correlations with $\sigma_{\rm e}$ here. It is the first time that the dark matter fraction variation is studied within such a wide velocity dispersion range ($1.6\lesssim \lg\sigma_{\rm e}\lesssim 2.6$) with accurate dynamical modelling in such a large sample.

\section{IMF variation}
\label{sec:imf}

\subsection{IMF from the $M/L-\sigma$ relations}

\begin{figure*}
	\centering
	\includegraphics[width=2\columnwidth]{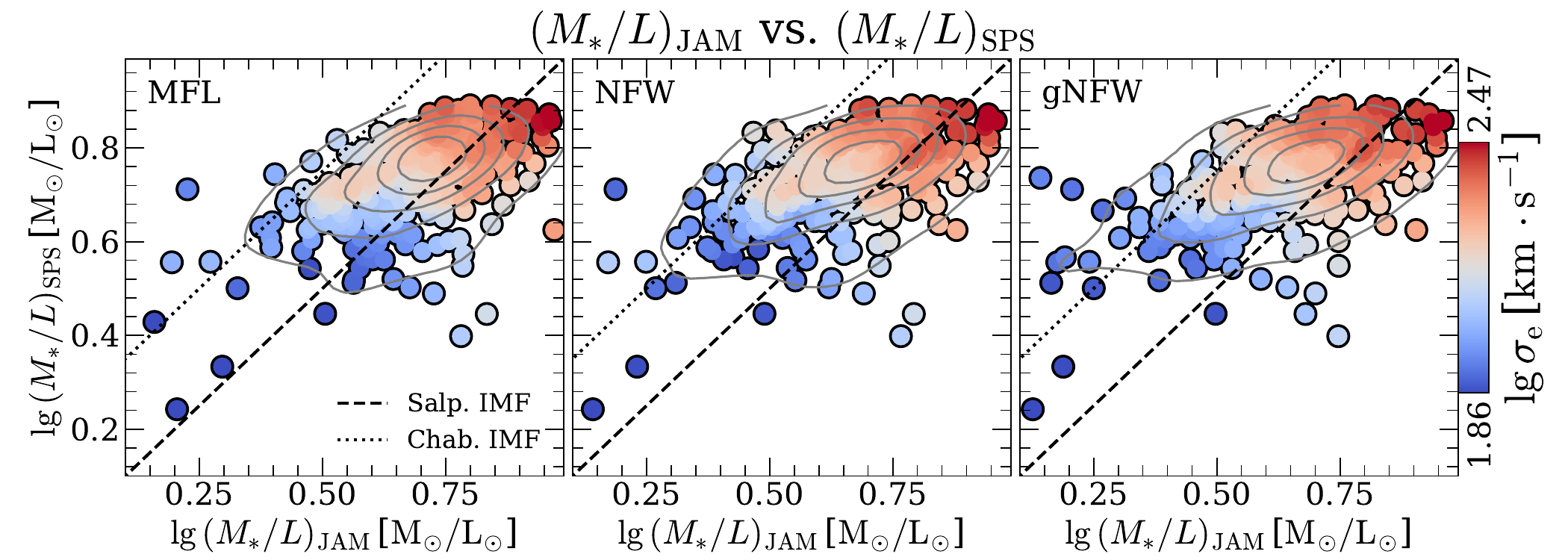}
	\caption{The comparison between the SPS-based stellar mass-to-light ratio and the JAM-based stellar mass-to-light ratio for the high quality ETG sample. The results from JAM with the MFL model (for this model, we take its total mass-to-light ratio instead of the stellar one, as we do not separate stars and dark matter in this model), the NFW model, and the gNFW model are shown from left to right. In each panel, the colours indicate the velocity dispersion of the galaxies. The black dotted line indicates the Chabrier IMF ($y=x+0.25$) and the black dashed line indicates the Salpeter IMF (adopted the SPS fitting; $y=x$). A kernel density estimation of the galaxy number density is indicated by the grey contours.}
	\label{fig:mlcompare}
\end{figure*}

\autoref{fig:fdm2sigma} already demonstrates, in qualitative manner and for a much larger sample, the main IMF trends that we will discuss more in detail later for our smaller golden sample: (i) the fact that for $\lg\,\sigma_{\rm e}\la2.3$, the median $f_{\rm DM}^{\rm SPS}$ becomes negative, indicates that the IMF must be lighter than the adopted Salpeter IMF at those lower $\sigma_{\rm e}$, consistently with previous studies; (ii) the average $f_{\rm DM}^{\rm JAM}$ measured by JAM, which is independent of the stellar population, is nearly constant around $f_{\rm DM}^{\rm JAM}\approx10\%$ above  $\lg\sigma_{\rm e}\ga2.1$, while the $f_{\rm DM}^{\rm SPS}$, assuming a Salpeter IMF, increases with $\sigma_{\rm e}$. Given that dark matter cannot explain the trend in $f_{\rm DM}^{\rm SPS}$, this suggests that the IMF becomes heavier with $\sigma_{\rm e}$. The variation $\Delta f_{\rm DM}^{\rm SPS}\approx0.2$ from the minimum is similar to the expected $M_{\ast}/L$ variation between a Salpeter and Chabrier IMF.

In \autoref{fig:mlcompare}, we present a direct comparison between the JAM-based and the SPS-based stellar mass-to-light ratios for the morphology-based sample (235 high-quality ETGs; see \autoref{sec:sample} for details of sample selection) for different JAM models (i.e. the MFL model\footnote{For the MFL model, we take its total mass-to-light ratio instead of stellar mass-to-light ratio, as we do not separate stars and dark matter in this model.}, the NFW model, and the gNFW model). As shown in the figure, for all the JAM models, galaxies with low stellar mass-to-light ratio typically have higher $(M_{\ast}/L)_{\rm SPS}$ than $(M_{\ast}/L)_{\rm JAM}$, consistent with \citet{Cappellari_et_al.(2012)}. Moreover, a similar trend is also seen for $\sigma_{\rm e}$, where we see galaxies with low $\sigma_{\rm e}$ have higher $(M_{\ast}/L)_{\rm SPS}$ than $(M_{\ast}/L)_{\rm JAM}$ (i.e. the assumed Salpeter IMF is heavier than reality). Regardless of its interpretation, the trend between the stellar population and dynamical stellar mass-to-light ratio is a robust and very general empirical result: \autoref{fig:mlcompare} can be compared directly with \citet[fig.~11]{Cappellari_et_al.(2013b)} for the ATLAS$^{\rm 3D}$ sample in the local Universe and with \citet[fig.~9]{Cappellari(2023)} at redshift $z\approx0.8$.

\subsection{IMF vs velocity dispersion}
\label{sec:imf2sigma}

In this section, we present the initial mass function (IMF) variation as a function of velocity dispersion of galaxies. To quantify the IMF and its variation, we follow the practice of \citet{Treu_et_al.(2010)}, and define the stellar mass excess factor (also known as IMF mismatch parameter) as:
\begin{equation}
	\label{eq:alpha}
	\alpha_{\rm IMF} \equiv \frac{(M_{\ast}/L)_{\rm JAM}}{(M_{\ast}/L)_{\rm SPS}},
\end{equation}
where $(M_{\ast}/L)_{\rm JAM}$ and $(M_{\ast}/L)_{\rm SPS}$ are the stellar mass-to-light ratio estimated with JAM and SPS, respectively (see \autoref{sec:jam} and \autoref{sec:sps} for details). Before making the comparison between the two stellar mass-to-light ratios, the JAM-based stellar mass-to-light ratio is corrected for the dust attenuation effect (both the dust influence from the Milky Way and the galaxy itself; see \autoref{sec:sps} for more details) in order to be comparable with the SPS-based stellar mass-to-light ratio. The SPS-based stellar mass-to-light ratio employed here is fitted with a \citet{Salpeter1955} IMF. Thus, $\alpha_{\rm IMF}=1$ (or equivalently, $\lg\,\alpha_{\rm IMF}=0$) indicates that JAM gives the same stellar mass estimate as SPS with a \citet{Salpeter1955} IMF\footnote{We will discuss more about this point in \autoref{sec:top_vs_bottom}.}.

\subsubsection{Comparison between JAM models}
\label{sec:imf2sigma_JAMmodels}

\begin{figure*}
    \centering
    \includegraphics[width=2\columnwidth]{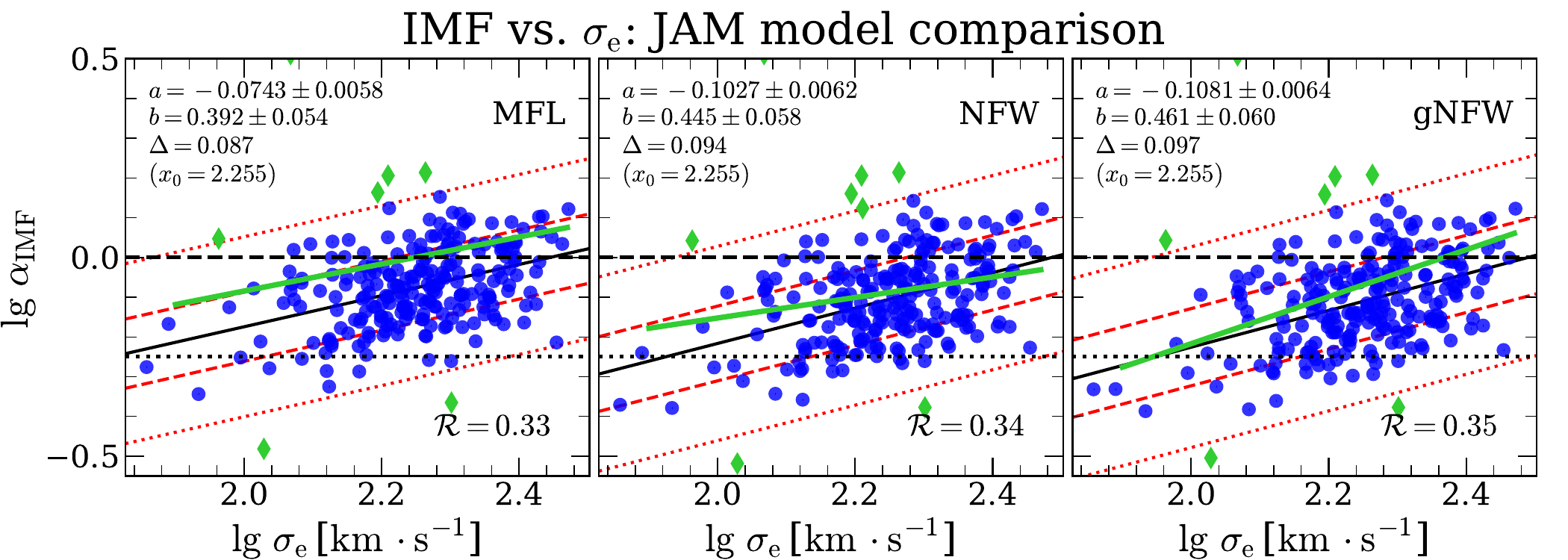}
    \caption{The correlation between $\alpha_{\rm IMF}$ ($\equiv (M_{\ast}/L)_{\rm JAM}/(M_{\ast}/L)_{\rm SPS}$) and $\sigma_{\rm e}$ of selected high-quality early-type galaxies for MFL, NFW, and gNFW models from left to right. In each panel, the blue dots are the galaxies used to fit the linear correlation and the green diamonds are the outliers excluded from the fit by {\sc LtsFit} procedure \citep{Cappellari_et_al.(2013a)}. The black solid line is the best-fitted $\lg\,\sigma_{\rm e}-\lg\,\alpha_{\rm IMF}$ obtained using the {\sc ppxf} \citep{Cappellari_et_al.(2013a)} software.  The red dashed lines and dotted lines indicate the 1$\sigma$ error (68 per cent) and 2.6$\sigma$ error (99 per cent), respectively. The green solid line is the best-fitted correlation between $\alpha_{\rm IMF}$ and $\sigma_{\rm e}$ from references with the same JAM mass model in each panel. The black dashed line and the black dotted line indicate the \citet{Salpeter1955} and the \citet{Chabrier(2003)} IMF. From left to right, the references are \citet{Shetty_et_al.(2020)}, \citet{Cappellari_et_al.(2013b)}, and \citet{Li_et_al.(2017)}. The best-fitted coefficients of the linear correlation $y=a+b(x-x_0)$ (where $y$ is $\lg\,\alpha_{\rm IMF}$ and $x$ is $\lg\,\sigma_{\rm e}$), the root-mean-square scatter $\Delta$, and the Pearson correlation coefficient $\mathcal{R}$ are listed in each panel.}
    \label{fig:imf2sigma_model}
\end{figure*}
    
In \autoref{fig:imf2sigma_model}, we present the correlation between the stellar mass excess factor $\alpha_{\rm IMF}$ and velocity dispersion $\sigma_{\rm e}$ for 3 different JAM models (MFL, NFW, and gNFW; see \autoref{sec:jam_models} and sec. 3.3 of \citetalias{Zhu_et_al.(2023a)} for more details of the mass models). The galaxies used here are strictly selected early-type galaxies, which have reliable JAM-based stellar mass-to-light ratio estimate (i.e. the morphology-based sample; see \autoref{sec:sample} for details of sample selection). All linear fittings in this paper are performed with the \textsc{LtsFit} package\footnote{Available from \url{https://pypi.org/project/ltsfit/}.} described in \citet{Cappellari_et_al.(2013a)}. This tool combines the Least Trimmed Squares robust technique of \citet{Rousseeuw_et_al.(2006)} into a least-squares fitting algorithm which can account for errors in all variables, intrinsic scatter, and automatic outlier detection. As can be seen, a clear positive correlation between $\alpha_{\rm IMF}$ and $\sigma_{\rm e}$ is seen for all the 3 JAM models, confirming that galaxies with low $\sigma_{\rm e}$ tend to have \citet{Chabrier(2003)} IMF (or even lighter), while high $\sigma_{\rm e}$ galaxies tend to have \citet{Salpeter1955} IMF (or even heavier), consistent with previous studies (e.g. \citealt{Cappellari_et_al.(2012),Cappellari_et_al.(2013b),Li_et_al.(2017),Shetty_et_al.(2020)}) with the same method. 

$\alpha_{\rm IMF}$ from MFL model is systematically higher than those from NFW and gNFW models, as we did not separate stars and dark matter in MFL model (i.e. the ``stellar'' mass-to-light ratio from MFL model is actually equivalent to the ``total'' mass-to-light ratio of galaxies). The JAM-based mass-to-light ratio from MFL sets the upper limit of stellar mass-to-light ratios from other JAM models. MFL model shows the flattest IMF$-\sigma_{\rm e}$ relation among the three investigated models. This is consistent with \autoref{fig:ml2sigma}, where we see that at low $\sigma_{\rm e}$ end the mass-to-light ratio from MFL model (i.e. the total mass-to-light ratio) is obviously higher than those from NFW model (i.e. dark matter is more important at low $\sigma_{\rm e}$; see also \autoref{fig:fdm2sigma}), while at high $\sigma_{\rm e}$ end the difference between MFL and NFW becomes smaller. This makes the stellar mass excess factors for MFL model is higher than that from stellar-DM decomposed models (i.e. NFW and gNFW model) at low $\sigma_{\rm e}$, while is similar at high $\sigma_{\rm e}$, making the $\sigma_{\rm e}-\alpha_{\rm IMF}$ trend for MFL model to be flatter. Interestingly, the IMF$-\sigma_{\rm e}$ relation under MFL model is slightly tighter than that under NFW/gNFW model, with the scatter being 0.087 dex (relative to 0.094 dex and 0.097 dex for NFW and gNFW models, respectively). This may be due to the stellar-DM degeneracy in dynamical modelling where the dynamical models always give only the {\em total} density, regardless of the modelling technique. The dark/luminous separation always depends on assumptions: the closer one assumes the shape of the stellar density traces the DM density, the strongest the degeneracy. This results in the more accurate estimate of {\em total} mass-to-light ratio (i.e. the mass-to-light ratio derived with an MFL model) than the stellar ones. The scatter discrepancy is also seen in \citet[][fig.~7]{Shetty_et_al.(2020)}, where the scatter of IMF$-\sigma_{\rm e}$ relation under self-consistent model (equivalent to the MFL model in this work) is also smaller than that under the model with an assumed dark halo. The observed scatter is comparable but slightly smaller than that $\Delta=0.11$ dex in the ATLAS$^{\rm 3D}$ models \citep{Cappellari_et_al.(2013b)}. This may be due to the more uncertain relative distances for low redshift galaxies or the more limited wavelength range of the ATLAS$^{\rm 3D}$ reducing the accuracy of the population $M_{\ast}/L$. The scatter is instead slightly larger than the one $\Delta=0.071$ for the Coma cluster study by \citet{Shetty_et_al.(2020)}, where relative distances are all accurately known and wavelength coverage is similar.

In each panel, we also plot the best-fitted results of previous studies with the same JAM mass model for comparison. In general, our results agree well with the results in previous studies within $1\sigma$ level and only show minor differences: $\alpha_{\rm IMF}$ of \citet{Shetty_et_al.(2020)} with MFL model is $\sim 0.07\,\rm dex$ higher than our result with similar slope. The results from \citet{Cappellari_et_al.(2013b)} (NFW model) and \citet{Li_et_al.(2017)} (gNFW model) show slightly flatter and steeper correlation between $\alpha_{\rm IMF}$ and $\sigma_{\rm e}$, respectively, compared to our results. This confirms the robustness of our estimate of the stellar mass excess factor and again confirms the trend between stellar population and dynamical stellar mass estimates as a function of stellar velocity dispersion. This has been interpreted as due to the IMF variation among different galaxies.

\subsubsection{Comparison between sample selection criteria}
\label{sec:comparison_sample}

\begin{figure*}
    \centering
    \includegraphics[width=2\columnwidth]{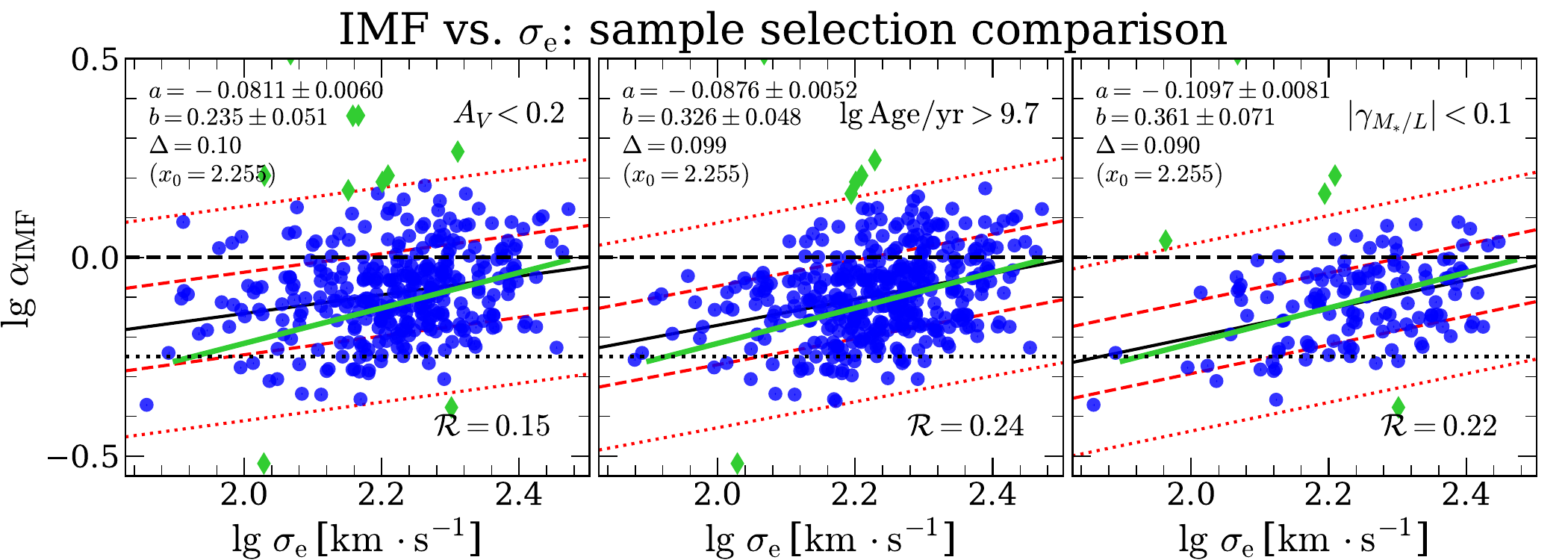}
    \caption{The correlation between $\alpha_{\rm IMF}$ and $\sigma_{\rm e}$ of the dust-based sample (left), the age-based sample (middle), and the $\gamma_{M_{\ast}/L}$-based sample (see \autoref{sec:comparison_sample} for definitions of the samples). In each panel, the green solid line is the best-fitted result of the high-quality ETG sample (with an NFW model). The other symbols are the same as \autoref{fig:imf2sigma_model}. The best-fitted coefficients of the linear correlation $y=a+b(x-x_0)$ (where $y$ is $\lg\,\alpha_{\rm IMF}$ and $x$ is $\lg\,\sigma_{\rm e}$), the root-mean-square scatter $\Delta$, and the Pearson correlation coefficient $\mathcal{R}$ are listed in each panel.}
    \label{fig:imf2sigma_sample}
\end{figure*}
        
In this section, we study the impact of sample selection criteria on the IMF$-\sigma_{\rm e}$ relation. In the previous section, we select the early-type galaxies for the IMF investigation due to the strong influence of dust and gas on stellar mass-to-light ratio in late-type galaxies (i.e. the morphology-based sample; see \autoref{sec:sample} for details). Actually, some other criteria can also be applied to reduce the negative influence of dust and gas in galaxies on the estimate of IMF, for example, selecting by age and dust attenuation factor. Thus, we further select two sub-samples from the golden sample (i.e. 909 galaxies with the most reliable estimates of JAM-based stellar mass-to-light ratio; see \autoref{sec:sample} for details):
\begin{itemize}
\item Dust-based sample: galaxies in this group are defined to be those with $A_V<0.2$ (where $A_V$ is the dust attenuation factor at $\lambda = 5500\,\Angstrom$, i.e. $V-$band), making up 340 galaxies among 909 galaxies in the golden sample.

\item Age-based sample: galaxies in this group are defined as the galaxies with $\lg\,\mathrm{Age}> 9.7$ in the golden sample, resulting in 394 galaxies out of 909 galaxies.
\end{itemize}

Besides, a constant stellar mass-to-light ratio is typically used in JAM, which, however, is not true under some circumstances (e.g. \citealt{Ge_et_al.(2021)}). \citetalias{Lu_et_al.(2023a)} pointed out that galaxies in green valley exhibit strongly decreasing stellar mass-to-light ratio profiles from galaxy centre to the outskirts using stellar population synthesis. \citetalias{Zhu_et_al.(2024)} also pointed out that the assumption of constant stellar mass-to-light ratio in JAM influences on the estimate of dark matter fraction (see sec. 3.4.1 therein). Thus here, we investigate the possible influence of the constant-$M_{\ast}/L$ assumption in JAM by selecting a sub-sample of galaxies with flat stellar mass-to-light ratio profiles on the morphology-based sample selected before:
\begin{itemize}
\item $\gamma_{M_{\ast}/L}$-based sample: galaxies in this group are the ones with $|\gamma_{M_{\ast}/L}|<0.1$, making up 133 galaxies among 235 high quality early-type galaxies in the morphology-based sample.
\end{itemize}

In \autoref{fig:imf2sigma_sample}, we present the $\lg\,\alpha_{\rm IMF}-\lg\,\sigma_{\rm e}$ relation for different samples described above. The JAM-based stellar mass-to-light ratio used here is under the NFW model (see \autoref{sec:jam} for details). As can be seen, the dust-based sample and age-based sample still show positive correlation between $\alpha_{\rm IMF}$ and $\sigma_{\rm e}$, while the relations seem to be not as tight as the relation for the morphology-based sample (i.e. larger scatters and smaller Pearson correlation coefficient). The slopes of $\lg\,\alpha_{\rm IMF}-\lg\,\sigma_{\rm e}$ relations with dust-based sample and age-based sample are both slightly smaller (i.e. flatter) than the morphology-based sample. Interestingly, the $\gamma_{M_{\ast}/L}$-based sample shows nearly the same $\lg\,\alpha_{\rm IMF}-\lg\,\sigma_{\rm e}$ relation as its parent sample (i.e., the morphology-based sample), with only the tightness slightly increasing. This is consistent with \citet{Li_et_al.(2017)}, which also found that the positive $\lg\,\alpha_{\rm IMF}-\lg\,\sigma_{\rm e}$ relation still exists when considering the stellar mass-to-light ratio gradient.

\subsection{JAM models with general $M_{\ast}/L$ gradients}
\label{sec:constant_ML_test}

\begin{figure*}
\centering
\includegraphics[width=1.8\columnwidth]{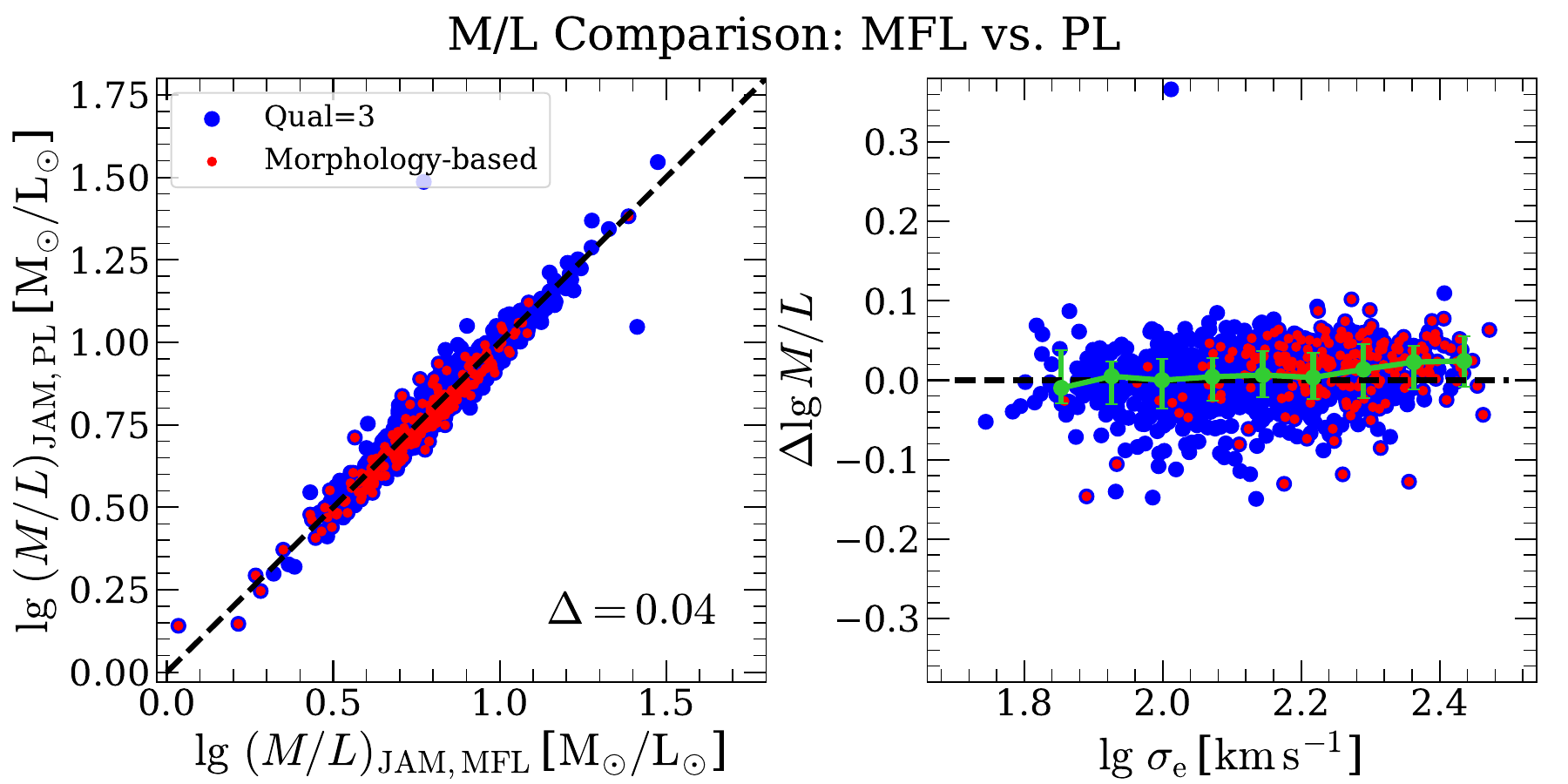}
\caption{Left: the comparison between the total mass-to-light ratios from the PL model and the MFL model (see \autoref{sec:jam_models} for details). The blue dots are the galaxies with $\mathrm{Qual}= 3$, and the red points denote the morphology-based sample, which is the default sample for IMF study in this work (see \autoref{sec:sample} for details). The black dashed line indicates the $y=x$ relation. Right: trend of $\Delta \lg M/L$ ($\equiv \lg (M/L)_{\rm JAM,MFL} - \lg (M/L)_{\rm JAM,PL}$) as a function of velocity dispersion, $\sigma_{\rm e}$. The black dashed line indicates $\Delta \lg M/L = 0$, and the green curve is the median trend for the $\rm Qual=3$ sample, with error bars indicating the range from 16th to 84th percentiles ($\pm 1\sigma$). These plots show that the total $M/L$ is the same for the MFL models and for model that allows for density general gradients (i.e. the PL model; see \autoref{sec:constant_ML_test} for more details).}
\label{fig:ml_comparison_pl}
\end{figure*}

\begin{figure}
\centering
\includegraphics[width=1\columnwidth]{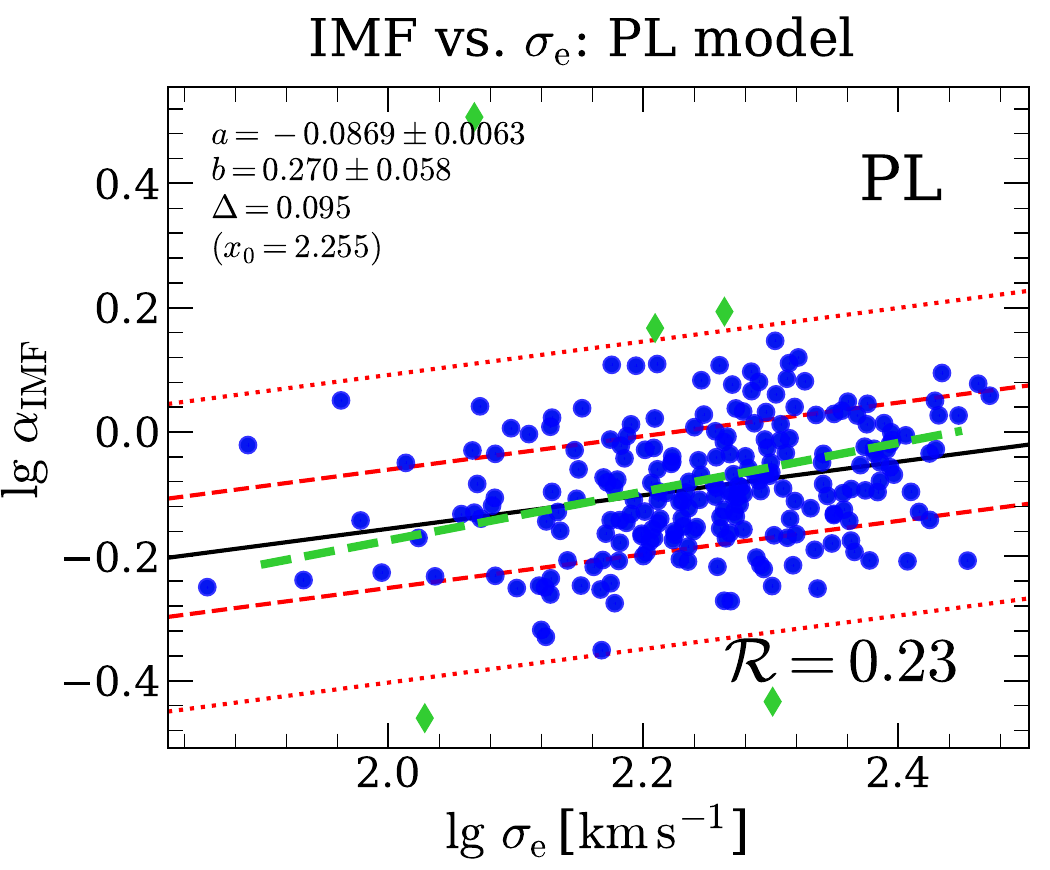}
\caption{Correlation between $\lg\,\alpha_{\rm IMF}$ (based on PL model) and $\lg\,\sigma_{\rm e}$. The green dashed line indicates the best fitting relation from MFL model. The other symbols are the same as \autoref{fig:imf2sigma_model}. The best-fitted coefficients of the linear correlation $y=a+b(x-x_0)$ (where $y$ is $\lg\,\alpha_{\rm IMF}$ and $x$ is $\lg\,\sigma_{\rm e}$), the root-mean-square scatter $\Delta$, and the Pearson correlation coefficient $\mathcal{R}$ are listed.}
\label{fig:alpha2sigma_pl}
\end{figure}

\begin{figure}
\centering
\includegraphics[width=1\columnwidth]{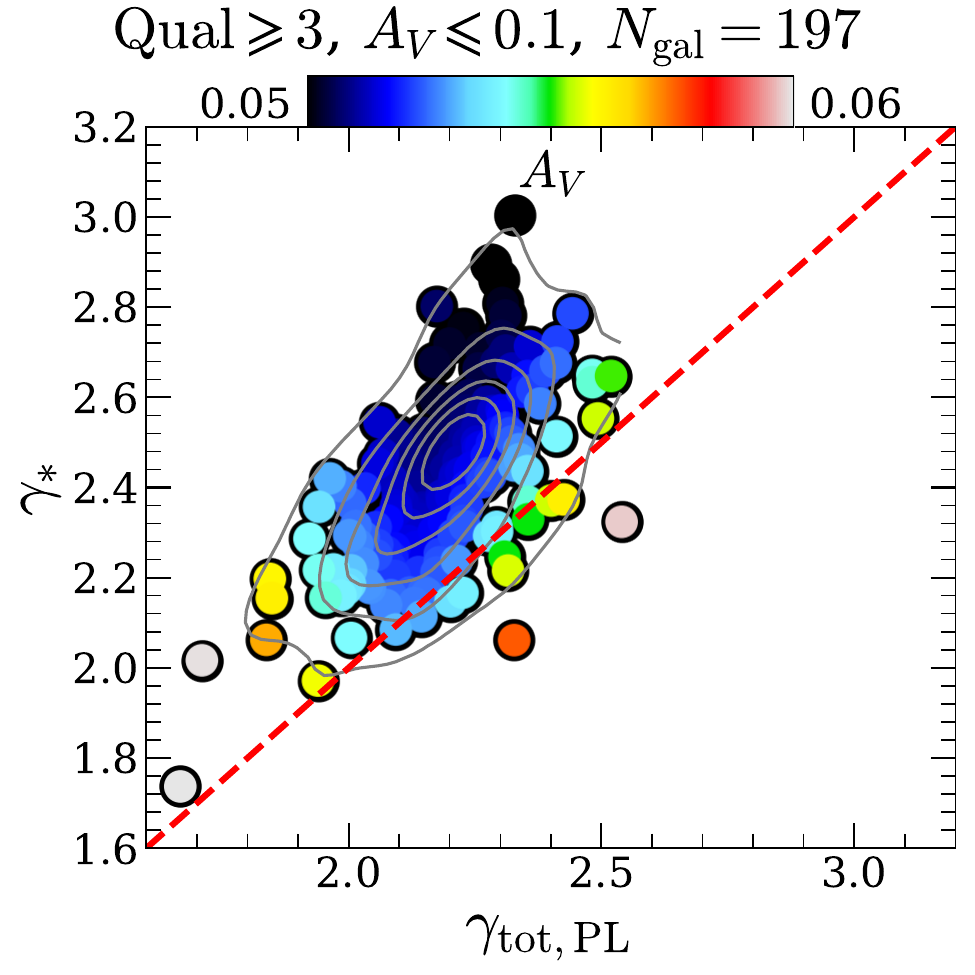}
\caption{The comparison between the total density slope from PL model ($\gamma_{\rm tot, PL}$) and the stellar density slope ($\gamma_{\ast}$; same as the luminosity density slope, as the stellar mass-to-light ratio is assumed to be constant) within $R_{\rm e}$ for galaxies with $\rm Qual\geqslant 3$ and $A_V\leqslant 0.1$ (see \autoref{sec:sps} for definition of $A_V$). The slopes are defined in \autoref{eq:density_slope}. The colours of the points indicate $A_V$. The red dashed line indicates $\gamma_{\rm tot, PL}=\gamma_{\ast}$. This plot indicates that the average \emph{total} $M/L$ decreases towards the centre within the studied region.}
\label{fig:gamma_comparison}
\end{figure}

We note that the stellar mass-to-light ratio gradients may be due to either the aging process and metallicity difference within individual galaxies (e.g. \citealt{Li_et_al.(2018),Ge_et_al.(2021)}; see also our \citetalias{Lu_et_al.(2023a)}) or the radial gradient of initial mass function (e.g. \citealt{van_Dokkum_et_al.(2017),Parikh_et_al.(2018),LaBarbera2019}), the latter of which may have even stronger impact on the estimate of IMF with the dynamical modelling-based method (e.g. \citealt{Bernardi_et_al.(2018),Bernardi_et_al.(2019),Bernardi_et_al.(2023a), Dominguez_et_al.(2019),Marsden_et_al.(2022),Mehrgan_et_al.(2024)}). Thus, in this section, we present more discussion of the effect of constant-$M_{\ast}/L$ assumption in JAM.

It has been argued that the presence of IMF gradients may bias the results of dynamical models that assume a constant IMF, and these gradients may artificially increase the observed IMF variations within $1R_{\rm e}$ in galaxies (e.g. \citealt{Bernardi_et_al.(2018),Dominguez_et_al.(2019)}). The bias due to IMF gradients was suggested as a possible explanation for the disagreement between global IMF within $1R_{\rm e}$ inferred from dynamics and population respectively \citep[see review by][]{Smith(2020)}. After the submission of our paper, similar claims were also made by \citet{Mehrgan_et_al.(2024)}.

To address these concerns, here we study the correlation between $\alpha_{\rm IMF}$ and $\sigma_{\rm e}$ while relaxing the constant stellar mass-to-light ratio assumption in the JAM models. For this, we build another JAM model in addition to the existing 4 mass models (i.e. MFL, NFW, gNFW, and fixed NFW models) in \citetalias{Zhu_et_al.(2023a)}. The new model does not distinguish the stellar and dark matter components, but only parametrizes the total density. It is identical to Model~(e) of \citet{Mitzkus2017} and Model~I of \citet{Poci2017} and is characterized by a \emph{total} density parametrized by a spherical generalized NFW profile \citep{Wyithe2001}:
\begin{equation}
    \rho_{\rm tot}(r)=\rho_{\rm s}\left(\frac{r}{r_{\rm s}}\right)^{\gamma_{\rm PL}}
        \left(\frac{1}{2}+\frac{1}{2}\frac{r}{r_{\rm s}}\right)^{-\gamma_{\rm PL}-3}.
\end{equation}
Same as \citetalias{Zhu_et_al.(2023a)}, the break radius $r_{\rm s}$ here is set to be $5\times r_{\rm max, bin}<r_{\rm s}<250$ kpc, where $r_{\rm max, bin}$ is the largest radius of the galaxy where can be observed, to avoid unrealistic small/large $r_s$ (see sec. 3.3.2 of \citetalias{Zhu_et_al.(2023a)} for more details). With such a large break radius, the total density profile is nearly a power-law within the region covered by the kinematics ($1.5-2.5R_{\rm e}$ for MaNGA). Thus, we call the new model power-law (PL) model hereafter. The stellar luminous tracer population is still described by the MGE parametrization of the observed surface brightness in the DynPop catalogue \citetalias{Zhu_et_al.(2023a)}.

The PL model does not provide the stellar and dark mass separately but only returns the total density profile. Having the total mass profile and the luminosity profile, we are able to derive the total mass-to-light ratio profile and the average mass-to-light ratio within $R_{\rm e}$. The PL model is analogous to the MFL model, for the fact that it also does not differentiate between luminous and dark matter. However, unlike the MFL model, the PL one is totally independent of any assumptions on the variation of the stellar mass-to-light ratios. The boundary of total density slope ($\gamma_{\rm PL}$) is set to range from -4 to 0, allowing for steeper total density slopes than the stellar ones (unlike our previous models). Thus, it is ideal to test for the effect of possible $M/L$ gradients on the recovered total $M/L$ and the corresponding inferred trends of IMF among galaxies. The catalogue based on the PL model (including the total density slopes and the mass-to-light ratios) has been added to the JAM catalogue of \citetalias{Zhu_et_al.(2023a)}. Readers can obtain the PL model catalogue through the journal website or access the full catalogue, consisting of five different JAM models, from the MaNGA DynPop website (\url{https://manga-dynpop.github.io}).

In \autoref{fig:ml_comparison_pl}, we show the comparison between the total mass-to-light ratios from the PL model, $(M/L)_{\rm JAM,PL}$ (which makes no assumption on the mass-to-light ratio gradients) and those from the MFL model, $(M/L)_{\rm JAM,MFL}$, (the mass-follows-light model, which assumes a constant mass-to-light ratio; see \citetalias{Zhu_et_al.(2023a)} for more details), both of which are calculated within $R_{\rm e}$. For the PL models, we computed the $(M/L)_{\rm JAM,PL}$ by analytically integrating, within a sphere of radius $R_{\rm e}$, the MGEs of the best-fitting JAM models, using the \texttt{mge\_radial\_mass} procedure of the \textsc{JamPy} package. As can be seen, the total mass-to-light ratios from the two models agree remarkably well with each other, without detectable bias and with a rms scatter of only 0.04 dex. In the right panel of \autoref{fig:ml_comparison_pl}, we show the correlation between the logarithmic difference $\Delta \lg(M/L) \equiv \lg(M/L)_{\rm JAM,MFL} - \lg(M/L)_{\rm JAM,PL}$ of the two total mass-to-light ratios and the velocity dispersion, $\sigma_{\rm e}$. As shown in the figure, $\Delta \lg(M/L)$ shows no obvious correlation with $\sigma_{\rm e}$. These results confirm that the constant mass-to-light ratio assumption used in JAM will not influence the estimate of the average mass-to-light ratio, at least within $R_{\rm e}$, within which we study the IMF variation among galaxies. This indicates that $M/L$ gradients cannot be the reason for the observed trends in $\alpha_{\rm IMF}$

In \autoref{fig:alpha2sigma_pl}, we show the correlation between $\alpha_{\rm IMF}$ based on PL model and $\sigma_{\rm e}$, and make comparisons with the MFL model. Again, we can see that the $\alpha_{\rm IMF}-\sigma_{\rm e}$ relation from PL model, which makes no assumption on mass-to-light ratios, agrees remarkably well with the result from MFL model, which assumes a constant mass-to-light ratio. This again confirms the robustness of the estimate of mass-to-light ratio within $R_{\rm e}$ under the constant mass-to-light ratio assumption. Note that the $\alpha_{\rm IMF}$ trends inferred with MFL or PL models only reflect the true IMF if dark matter is negligible. But we showed in this paper that this is indeed the case for the studied galaxies and that the $\alpha_{\rm IMF}$ trends are the same when considering the NFW or gNFW models (\autoref{fig:imf2sigma_model}).

Our result contrasts with previous claims that $M_{\ast}/L$ gradients strongly affect the inferred total $M/L$ and may consequently spuriously drive the inferred IMF variations among galaxies \citep{Bernardi_et_al.(2018),Dominguez_et_al.(2019)}. The reason for this disagreement is unclear and would require further investigation. However, both previous papers use an approximate approach to estimate the influence of IMF gradients, while here we model the galaxy dynamics in detail for every galaxy. This may be the reason for the differences.

More recently, \citet{Mehrgan_et_al.(2024)} raised a similar concern as \citet{Bernardi_et_al.(2018)} and \citet{Dominguez_et_al.(2019)}, and pointed out that the constant-$(M_{\ast}/L)$ assumption will increase the estimate of stellar mass by up to a factor of 1.5. They found that the model allowing for mass-to-light ratio gradient predicts lower stellar mass excess factor compared to the model with constant stellar mass-to-light ratio (see their figs. 8 and 9). However, their study only considered 7 galaxies. In addition, they calculated the integrated stellar mass excess factor $\alpha$ as the light-weighted $\alpha (r)$ across different radii, which is also different from the definition in our work (see \autoref{eq:alpha}). They found that the IMF inferred from their dynamical models becomes significantly heavier, and the total $M/L$ increases towards the centre, within a region of radius $R\lesssim1$ kpc.

To assess for the presence of gradients in the total $M/L$ in our galaxies, in \autoref{fig:gamma_comparison} we compare the total density slopes from the PL model $\gamma_{\rm tot,PL}$ against the stellar density slopes (same as the luminosity density slope when stellar mass-to-light ratio is assumed to be constant) $\gamma_{\ast}$. The (total/stellar) density slopes are defined as:
\begin{equation}
\label{eq:density_slope}
\gamma = \frac{1}{\lg (R_{\rm e}/R_{\rm in})} \int_{R_{\rm in}}^{R_{\rm e}} \frac{\mathrm{d}\lg \rho}{\mathrm{d}\lg r} \mathrm{d}\lg r = \frac{\lg \rho(R_{\rm e})-\lg \rho(R_{\rm in})}{\lg R_{\rm e}-\lg R_{\rm in}},
\end{equation}
where $R_{\rm in}$ is the maximum between $0.1R_{\rm e}$ and the FWHM of MaNGA PSF; $\gamma$ and $\rho$ represent the total or stellar density slopes and radial densities.

Here, we select a subsample with the best dynamical models (i.e. $\mathrm{Qual}\geqslant3$) and negligible dust extinction $A_V\leqslant0.1$. We find that the total density slope is systematically smaller than the luminosity density. In other words, the total $M/L$ decreases towards the centre within $1R_{\rm e}$. This decrease in $M/L$ can be explained in two ways: (i) if dark matter is negligible, the decrease of the total $M/L$ implies a decrease of the stellar $M_{\ast}/L$, or (ii) if dark matter dominates, the gradient would be due to an increase of dark matter with increasing radius. Given that several studies, including this paper, indicate that dark matter is likely negligible within $1R_{\rm e}$ in ETGs, option (i) seems favored. 

This may seem in contrast with results indicating IMF becoming heavier in galaxy centres. However, if the IMF becomes heavier only within a very small radius $R\lesssim1$ kpc as suggested by \citet{Mehrgan_et_al.(2024)}, the effect would be undetectable by MaNGA, given that the relevant radius is generally contained within the FWHM of the PSF of the observations. Moreover, our power-law approximation for the total density is likely to break down at such small scales, and not provide a good approximation for the total density profile. In other words, we cannot confirm or disprove the possible presence of a heavier IMF within $R\lesssim 1$ kpc.
 
To summarize, we find that the global trends with $\sigma$ in $\alpha_{\rm IMF}$ that we infer from our models in \autoref{sec:imf2sigma} remain the same regardless of whether we allow for gradients in the total $M/L$ (or IMF) or not.

\subsection{IMF vs. stellar population properties}
In previous sections, we already see that the stellar mass excess factor, which constrains the IMF shape, shows positive correlation with velocity dispersion of galaxies. However, we still do not know whether $\sigma_{\rm e}$ is the main driver of IMF variation. In this section, we present the correlation between the stellar mass excess factor, $\alpha_{\rm IMF}$ and stellar population properties (including luminosity-weighted age, metallicity, and SPS-based stellar mass-to-light ratio; obtained from \citetalias{Lu_et_al.(2023a)}; see \autoref{sec:sps} for details) on the morphology-based sample (see \autoref{sec:sample} for selection details). 

\begin{figure*}
\centering
\includegraphics[width=2\columnwidth]{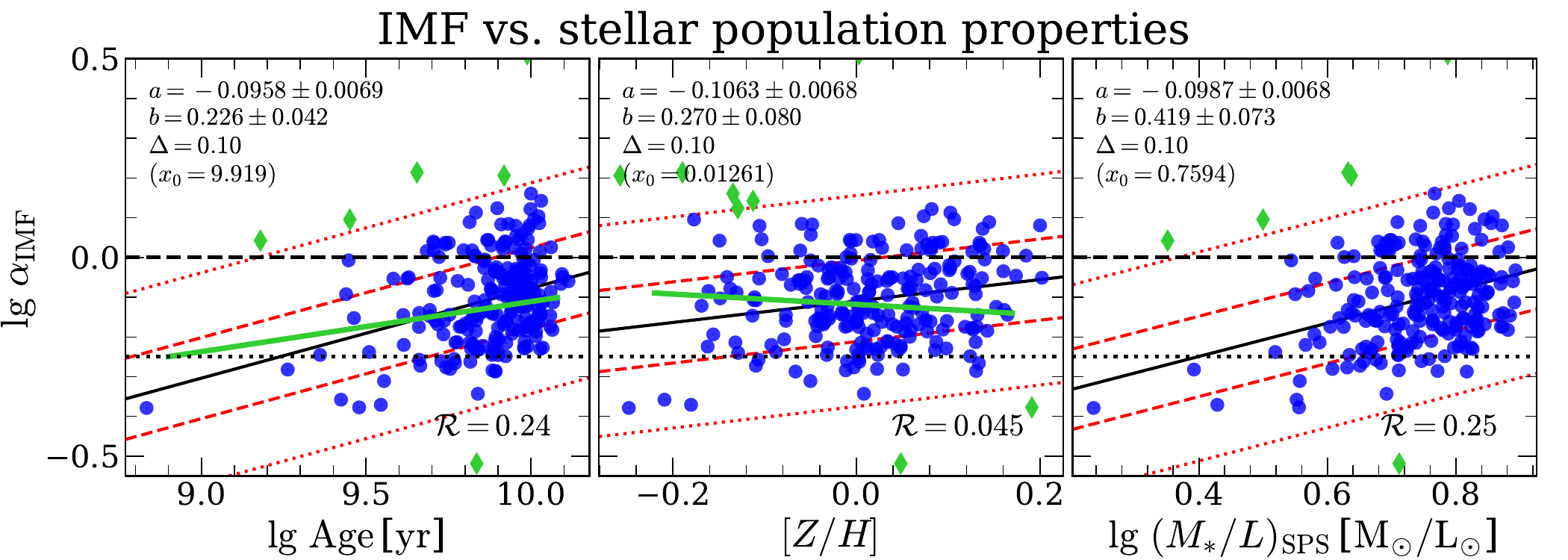}
\caption{The correlation between $\alpha_{\rm IMF}$ and stellar population properties (age, metallicity, and stellar mass-to-light ratio from left to right) for the high-quality ETG sample (under the NFW mass model). In the left and middle panels, the green solid lines are the results from \citet{McDermid_et_al.(2014)}. The Pearson correlation coefficient is listed in each panel.}
\label{fig:imf2sp}
\end{figure*}

In \autoref{fig:imf2sp}, we present the correlation between $\alpha_{\rm IMF}$ and age, metallicity, and SPS-based stellar mass-to-light ratio from left to right. As can be seen, $\alpha_{\rm IMF}$ show positive correlation with $\lg\,\mathrm{Age}$, consistent with \citet{McDermid_et_al.(2014)}, with the slope slightly larger (steeper) than that of \citet{McDermid_et_al.(2014)}. Metallicity, again similar to \citet{McDermid_et_al.(2014)}, does not show obvious correlation with $\alpha_{\rm IMF}$ (low Pearson correlation coefficient). A positive correlation between $\alpha_{\rm IMF}$ and SPS-based stellar mass-to-light ratio is also seen for the galaxies. Compared to these correlations, the IMF$-\sigma_{\rm e}$ relation is obviously stronger and tighter, indicating that $\sigma_{\rm e}$ is more likely the main driving factor of IMF variation among galaxies, compared to stellar population properties. 

We note, however, this is inconsistent with the studies which made use of the gravity-sensitive absorption lines, e.g. $\mathrm{Na_{I}}$ doublet and $\mathrm{FeH}$ (also known as the Wing-Ford band) or the full spectrum fitting to study the IMF shape as a function of metallicity (e.g. \citealt{Conroy_et_al.(2012),Martin-Navarro_et_al.(2015b),Zhou_et_al.(2019),Gu_et_al.(2022)}). They found an obvious correlation between IMF shape and metallicity, which may be even more responsible for the IMF variation than $\sigma_{\rm e}$ (e.g. \citealt{Zhou_et_al.(2019)}). \citet{Smith(2014)} studied this discrepancy between different methods with the stellar mass excess factors derived from dynamical modelling and spectrum analyses on the same sample. They found that although the stellar mass-to-light ratios derived in different methods both show positive correlation with velocity dispersion, there is no correlation between $\alpha_{\rm IMF}$ inferred from the two approaches on a galaxy-by-galaxy basis. They explained this as that one (or both) of the methods has not accounted fully for the main confounding factors and there is additional variation in the detailed shape of the IMF which cannot currently be inferred from the spectroscopic features. This remains a major crucial and worrying inconsistency between the different approaches. We will discuss it later in \autoref{sec:imf2metal_origin}.

\label{sec:imf2sp}

\section{IMF-sensitive spectral features}
\label{sec:spectra}
\begin{figure*}
\centering
\includegraphics[width=2\columnwidth]{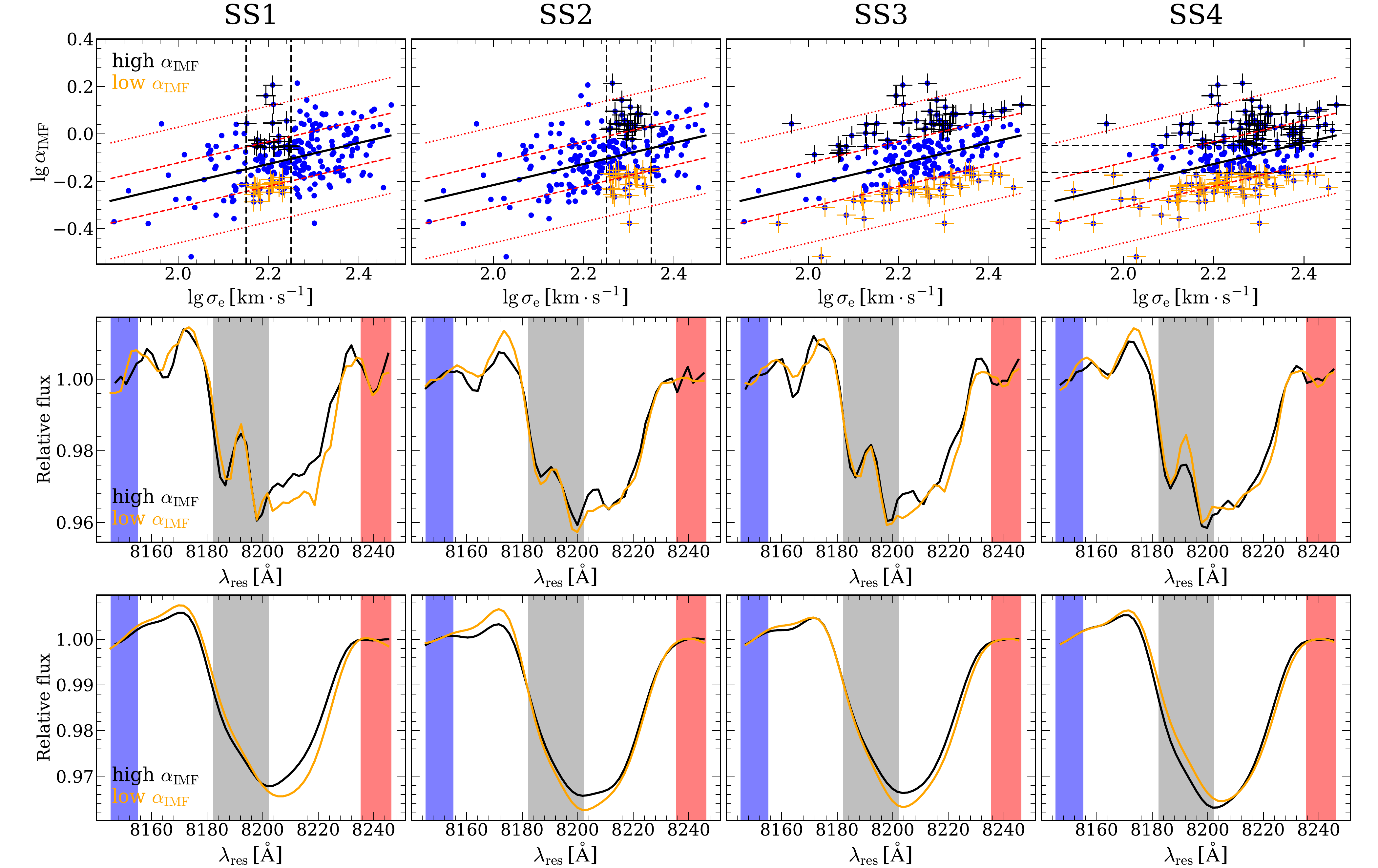}
\caption{Top: The sample selections of SS1-SS4 from left to right (see \autoref{sec:spectra} for definitions of the samples). In each panel, the parent sample is indicated by the blue dots (high-quality ETG sample, with an NFW mass model). The galaxies selected as high $\alpha_{\rm IMF}$ and low $\alpha_{\rm IMF}$ galaxies are indicated by black and orange plus symbols, respectively. In the first and second panel of the top row, the dashed vertical lines indicate the $\sigma_{\rm e}$ range, in which the samples are selected. In the fourth panel of the first row, the dashed horizontal lines indicate the lower (upper) $\alpha_{\rm IMF}$ boundary of the high (low) $\alpha_{\rm IMF}$ sample. Middle: The comparison of $\mathrm{Na_I}$ absorption line between the stacked spectra with high (black) and low (orange) $\alpha_{\rm IMF}$. Before plotting, the absorption feature is corrected for the local continuum (see \autoref{sec:spectra} for the definition of the local continuum). In each panel, the red and blue regions are the pseudo-continuum regions defined in \citet{La_Barbera_et_al.(2013)} and the grey band is the region of the feature. Bottom: The same as the middle row, while the spectra are smoothed to $250\,\mathrm{km\cdot s^{-1}}$ for visualization purpose.}
\label{fig:spec_part1}
\end{figure*}

\begin{figure}
\centering
\includegraphics[width=1\columnwidth]{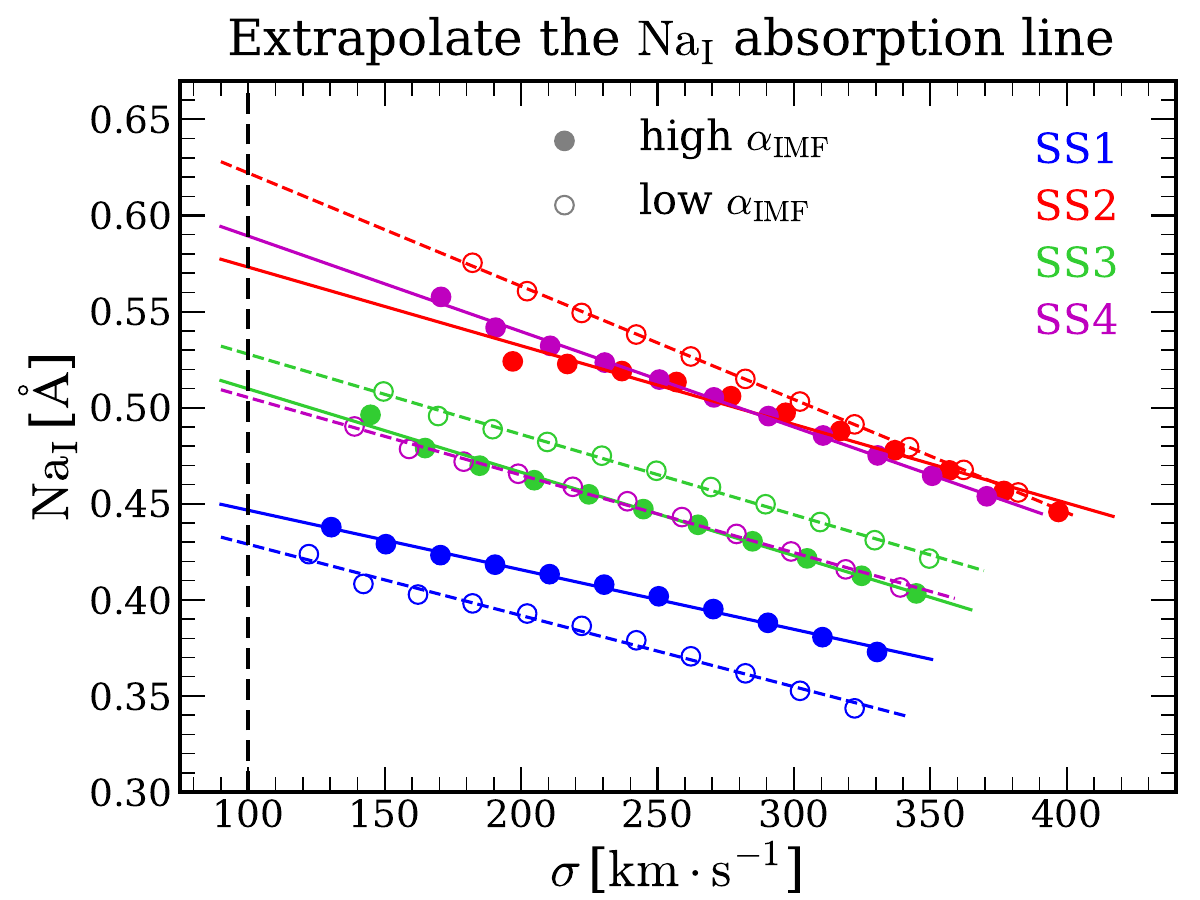}
\caption{The illustration of correcting the equivalent width of $\mathrm{Na_I}$ to $\sigma = 100\,\mathrm{km\cdot s^{-1}}$. Results for different samples (SS1-SS4; see \autoref{sec:spectra} for details of sample selection) are indicated by different colours. Filled and open circles denote the results of the spectrum with high and low $\alpha_{\rm IMF}$, respectively. The solid and dashed lines indicate the corresponding best-fitted $\mathrm{Na_I}-\sigma$ linear relations, fitted with the {\sc LtsFit} \citep{Cappellari_et_al.(2013a)} software. The black dashed line indicate $\sigma = 100\,\mathrm{km\cdot s^{-1}}$.}
\label{fig:extrapolate}
\end{figure}

\begin{figure*}
\centering
\includegraphics[width=2\columnwidth]{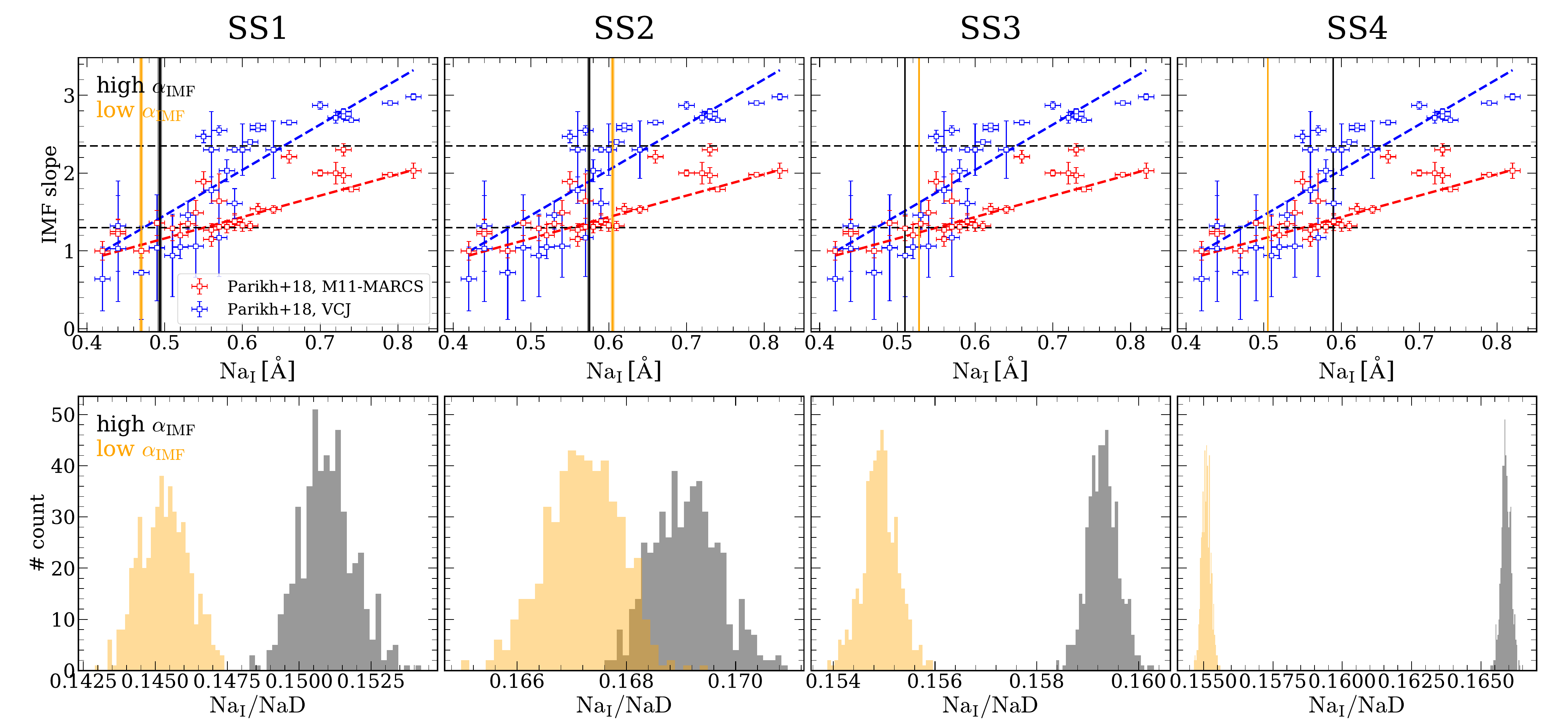}
\caption{Top: Correlation between IMF slope (within the mass range $0.1-0.5\,\mathrm{M_{\odot}}$) and the corrected $\mathrm{Na_I}$ equivalent widths (at $\sigma = 100\,\mathrm{km\cdot s^{-1}}$). The reference values are taken from \citet{Parikh_et_al.(2018)} and the red and blue data points indicate the results from different methods (see \citealt{Parikh_et_al.(2018)} for details). In each panel, the blue and red dashed lines indicate the best-fitted linear relation between IMF slope and $\mathrm{Na_I}$ equivalent width, obtained with \textsc{LtsFit} software \citep{Cappellari_et_al.(2013a)}. The black and orange vertical lines refer to the $\mathrm{Na_I}$ equivalent widths of our stacked spectra, with the shaded region indicating the $1\sigma$ region obtained with a Monte Carlo-base method (see \autoref{sec:spectra} for details). The \citet{Salpeter1955} IMF and the \citet{Kroupa(2001)} IMF are shown with the horizontal black dashed lines from top to bottom. Bottom: the distributions of the $\rm Na_I/NaD$ ratio for the high $\alpha_{\rm IMF}$ galaxies (black) and low $\alpha_{\rm IMF}$ galaxies (orange). For each stacked spectra, the distribution of $\rm Na_I/NaD$ is obtained through a Monte Carlo-base method (see \autoref{sec:spectra} for details) and both features are corrected to $\sigma = 100\,\mathrm{km\cdot s^{-1}}$.}
\label{fig:spec_part2}
\end{figure*}

\citet{Smith(2020)} suggested that one possibility to explain the inconsistent stellar mass excess factors derived with dynamical modelling and spectrum analysis is the aperture difference in which the stellar mass excess factors are measured, combined with radial gradients in the IMF: the dynamical modelling-based one is an average value within a sphere of approximately $1R_{\rm e}$, while the spectrum-based one is measured within $R_{\rm e}/8$. Thus, in this section, we combine the dynamical properties from \citetalias{Zhu_et_al.(2023a)} and the stellar population properties from \citetalias{Lu_et_al.(2023a)}, which are, by design, measured in the same aperture (i.e. the elliptical half-light isophotes), to study whether galaxies with high and low dynamical modelling-based $\alpha_{\rm IMF}$ show corresponding spectral differences. It is worth remembering that formally, the dynamics infers average $M/L$ inside a sphere of radius $R_{\rm e}$, while the stellar population measures average $M/L$ inside a projected (elliptical) cylinder of radius $R_{\rm e}$. In our comparison we ignore the likely small bias introduced by stars falling within the cylinder but outside the sphere of radius $R_{\rm e}$.

To make the investigation, we first select 4 sub-samples from 235 high quality early-type galaxies in the morphology-based sample (see \autoref{sec:sample} for details). Each of them contains two sub-groups with different dynamical $\alpha_{\rm IMF}$, with the criteria being:
\begin{enumerate}
\item Sub-sample 1 (SS1 hereafter): galaxies in this sub-sample are defined to have similar $\sigma_{\rm e}$, i.e. $2.15<\lg\,(\sigma_{\rm e}/\mathrm{km\cdot s^{-1}})<2.25$. Then the high-$\alpha_{\rm IMF}$ (low-$\alpha_{\rm IMF}$) sub-group is further defined as those with the highest (lowest) 30\% $\alpha_{\rm IMF}$ within the $\sigma_{\rm e}$ bin, resulting in 21 galaxies in each sub-group.

\item Sub-sample 2 (SS2 hereafter): galaxies in this sub-sample are similar with SS1, but within a $\sigma_{\rm e}$ range from $\lg\,(\sigma_{\rm e}/\mathrm{km\cdot s^{-1}})=2.25$ to $\lg\,(\sigma_{\rm e}/\mathrm{km\cdot s^{-1}})=2.35$, resulting in 25 galaxies in both the sub-group with high $\alpha_{\rm IMF}$ and the one with low $\alpha_{\rm IMF}$.

\item Sub-sample 3 (SS3 hereafter): the high-$\alpha_{\rm IMF}$ (low-$\alpha_{\rm IMF}$) sub-group in this sub-sample contains the galaxies whose $\alpha_{\rm IMF}$ is $1\sigma$ higher (lower) than the best-fitted $\alpha_{\rm IMF}-\sigma_{\rm e}$ correlation, resulting in 47 galaxies with high $\alpha_{\rm IMF}$ and 43 galaxies with low $\alpha_{\rm IMF}$.

\item Sub-sample 4 (SS4 hereafter): the high-$\alpha_{\rm IMF}$ (low-$\alpha_{\rm IMF}$) sub-group in this sub-sample contains the galaxies whose $\alpha_{\rm IMF}$ is the highest (lowest) 30\% among all the galaxies, regardless of $\sigma_{\rm e}$, resulting in 71 galaxies with high and low $\alpha_{\rm IMF}$.
\end{enumerate}

To better demonstrate the sample selection, we present, in \autoref{fig:spec_part1} (the top panels), the distributions of the selected galaxies on the $(\sigma_{\rm e},\alpha_{\rm IMF})$ plane for the 4 sub-samples. The relevant information of these 4 sub-samples are also listed in \autoref{table:spec_samples}. Having the samples, we then stack the spectra of the galaxies in the same sub-group for further investigation following the practice of \citet{Parikh_et_al.(2018)}. The steps are:
\begin{enumerate}
\item For a single galaxy, the spectrum in each spaxel is corrected to the rest frame with the corresponding redshift from NASA Sloan Atlas (NSA) catalogue \citep{Blanton_et_al.(2011)} and the corresponding line-of-sight velocity provided by the MaNGA Data Analysis Pipeline (DAP; \citealt{Belfiore_et_al.(2019),Westfall_et_al.(2019)}), using:
\begin{equation}
\lambda_{\rm res} = \frac{\lambda}{(1+z)\cdot (1+\frac{v}{c})},
\end{equation}
where $\lambda$ is the uncorrected (i.e. observed) wavelength; $z$ is the redshift of the galaxy; $v$ is the line-of-sight velocity of the spaxel\footnote{The velocities are obtained from the files under \texttt{manga-[PLATE]-[IFUDESIGN]-MAPS-SPX-MILESHC-MASTARSSP.fits}. We note that the velocity estimates of the pixel at the outskirts of the galaxies would be inaccurate without being binned to higher signal-to-noise ratio. Here, we only stack the spectra within $1R_{\rm e}$ and as a result, this effect would not largely change our results.}; $c$ is the light speed. The spectra used here are from the MaNGA DRP \citep{Law_et_al.(2016)} (under the names being \texttt{manga-[PLATE]-[IFUDESIGN]-LOGCUBE.fits}).

\item The spectrum of each spaxel is interpolated to a common wavelength grid, keeping the spectrum resolution unchanged.

\item Spectra within the elliptical half-light isophote of one individual galaxy are simply co-added, without normalization. This produces one stacked `effective spectrum' per galaxy.

\item The effective spectra of different galaxies are normalized with the mean flux within the wavelength range $6780\Angstrom - 6867\Angstrom$, in order to account for the variation in fluxes from galaxy to galaxy. After that, the normalized effective spectra of the galaxies within the same sub-group are further co-added.
\end{enumerate}

The stacked spectra so derived wash out the contamination of sky lines in the near-infrared (NIR) region of the spectrum, where the IMF-sensitive spectral features (i.e. $\mathrm{Na_I}$ and $\mathrm{FeH}$) exist. Here, we define the noise of the stacked spectra as $\sigma/\sqrt{N_{\rm gal}}$, where $\sigma$ is the standard deviation of the normalized flux of the spectra for stacking at specific wavelength and $N_{\rm gal}$ is the number of the galaxies stacked together. \citet{Parikh_et_al.(2018)} tried to use both the $\mathrm{Na_I}$ doublet and the $\mathrm{FeH}$ line to constrain the IMF of MaNGA galaxies but they found that the results based on the $\mathrm{FeH}$ line have large uncertainties. They argued that it is because the $\mathrm{FeH}$ feature is faint, compared to the bright sky lines in this region, making the small uncertainties in sky subtraction unable to cancel out in the stacked spectra. Thus, in this work, we only use the $\mathrm{Na_I}$ doublet to constrain the IMF shape in this study. As mentioned in \citet{van_Dokkum_et_al.(2012)}, $\mathrm{Na_I}$ line for a fixed Na elemental abundance is only present in low mass stars, and hence is able to constrain the richness of the low mass stars: the stronger the $\mathrm{Na_I}$ line, the more bottom heavy the IMF is. The wavelength range of the $\mathrm{Na_I}$ doublet, as well as the corresponding blue and red pseudo-continuum bandpasses are defined in \citet{La_Barbera_et_al.(2013)}. In \autoref{table:spec_samples}, we list the average $S/N$ of $\mathrm{Na_I}$ doublet for different sub-groups. As can be seen, with such stacking processes, the signal-to-noise ratio of the stacked spectra reaches a significantly high value, which allows for the spectral study of IMF.

We note here that in this work, we do not attempt to make detailed analysis on the IMF-sensitive spectral features, as done in previous studies \citep[e.g.][]{van_Dokkum_et_al.(2012),Conroy_et_al.(2012),Spiniello2012,Spiniello2015,La_Barbera_et_al.(2013),Parikh_et_al.(2018)}. We only want to see whether the galaxies with different dynamical $\alpha_{\rm IMF}$ show various IMF-sensitive spectral features, which confirm their IMF difference in a model-independent manner. In the middle panels of \autoref{fig:spec_part1}, we present the comparison of $\mathrm{Na_I}$ doublets between the stacked high $\alpha_{\rm IMF}$ spectrum and the low $\alpha_{\rm IMF}$ spectrum for the 4 different sub-samples. Before over-plotting, we first correct for the continuum shape of the spectrum and extract the absorption feature. The continuum is defined following the practice of \citet{Worthey(1994a)}: We first calculate the average flux in the blue and red pseudo-continuum bandpasses, using eq. (2) of \citet{Worthey(1994a)}. The local continuum is then defined by taking a straight line from the blue pseudo-continuum to the red pseudo-continuum. The relative flux of the absorption feature is calculated as the ratio between the true flux at a given wavelength and the flux of the local continuum at the same wavelength. For visualization, we further smooth the $\mathrm{Na_I}$ lines to $\sigma = 250\,\mathrm{km\cdot s^{-1}}$ in the bottom panels of \autoref{fig:spec_part1}. As can be seen, all the stacked spectra show clear $\mathrm{Na_I}$ absorption features and the doublet is washed out at high $\sigma$. Comparing the $\mathrm{Na_I}$ absorption features of galaxies in the high $\alpha_{\rm IMF}$ sub-group (black curves) and in the low $\alpha_{\rm IMF}$ sub-group (orange curves) in the same sub-sample, we find that the difference between the two spectra are quite small and highly depend on the sample selection. The stacked spectrum with high $\alpha_{\rm IMF}$ (black) of SS1 shows slightly stronger $\mathrm{Na_I}$ absorption feature than the low $\alpha_{\rm IMF}$ (orange). The difference is more significant in SS4, where more galaxies are stacked and a higher $S/N$ is reached. In SS2 and SS3, however, we see the opposite trend, where the stacked spectra with low $\alpha_{\rm IMF}$ (orange) show stronger $\mathrm{Na_I}$ absorption signals than the high $\alpha_{\rm IMF}$ ones. For sure, the variation we see is not comparable to the one illustrated fig.~1 of \citet{van_Dokkum_et_al.(2010)}. If the dynamically-determined $\alpha_{\rm IMF}$ variations are due to an IMF variation, the spectral features excludes the possibility for the IMF to vary from Chabrier to even heavier than Salpeter, as a number of studies originally suggested.

To quantitatively describe the difference of $\mathrm{Na_I}$ absorption strength, we calculate the equivalent width of the $\mathrm{Na_I}$ absorption feature with the method described in \citet{Worthey(1994a)} (see eqs. 2 and 3 therein). To build up the connection between $\mathrm{Na_I}$ equivalent width and the IMF shape, we take the values from \citet{Parikh_et_al.(2018)} (tables B1-B3) as references. The estimate of the equivalent widths of absorption lines always require a correction to homogenize all spectra to the same stellar velocity dispersion \citep[e.g.][]{Kuntschner(2004)}. Thus, to make the comparison between the $\mathrm{Na_I}$ equivalent widths of our spectra and those from \citet{Parikh_et_al.(2018)}, which are calculated at $\sigma = 100\,\mathrm{km\cdot s^{-1}}$, we have to first correct our $\mathrm{Na_I}$ equivalent widths to the same dispersion. \citet{Parikh_et_al.(2018)} (also \citealt{Westfall_et_al.(2019)}) convolved the best-fitted templates to obtain the $\mathrm{Na_I}$ equivalent width at the given velocity dispersion, while the templates used in this work do not have such high resolution, making it impossible to correct the equivalent widths with a simple convolution process. Here, we correct the $\mathrm{Na_I}$ equivalent widths following the steps below:
\begin{enumerate}
\item For each stacked spectrum, we estimate its dispersion ($\sigma_0$; typically larger than $100\,\mathrm{km\cdot s^{-1}}$) using {\sc ppxf}, allowing both the additive and multiplicative polynomials by setting \texttt{degree=mdegree=4}.

\item We smooth the stacked spectrum to larger dispersions and calculate the corresponding $\mathrm{Na_I}$ equivalent widths at every $20\,\mathrm{km\cdot s^{-1}}$. 

\item Finally, we perform a linear fit to the correlation using the \textsc{LtsFit} software \citep{Cappellari_et_al.(2013a)} and calculate the predicted $\mathrm{Na_I}$ equivalent width at $\sigma = 100\,\mathrm{km\cdot s^{-1}}$.

\item To derive the measurement uncertainty of the corrected $\mathrm{Na_I}$ equivalent widths, we employ a Monte Carlo-based method, following the practice of \citet{Parikh_et_al.(2018)}. Specifically, we perturb the flux at each wavelength with a number randomly taken from a Gaussian with its standard deviation equaling to the error in the flux at each pixel. Then the equivalent width is calculated and corrected to $\sigma = 100\,\mathrm{km\cdot s^{-1}}$. This is done for 500 times and the measurement uncertainty is measured as the standard deviation of the 500 $\mathrm{Na_I}$ equivalent widths.
\end{enumerate}

The correlation between $\sigma$ and the corresponding $\mathrm{Na_I}$ equivalent widths is shown in \autoref{fig:extrapolate}. As can be seen, this correlation can be well described by a linear relation, indicating the robustness of our extrapolation. In the top panels of \autoref{fig:spec_part2}, we over-plot the reference values of $\mathrm{Na_I}$ equivalent widths from \citet{Parikh_et_al.(2018)} and the corresponding IMF slopes against the corrected $\mathrm{Na_I}$ equivalent widths of this work. As can be seen, IMF slopes show positive correlation with $\mathrm{Na_I}$ equivalent width for both models adopted in \citet{Parikh_et_al.(2018)} (red and blue symbols; see sec.~2.4 therein for details of the two models). For SS1, we find that both the high $\alpha_{\rm IMF}$ spectrum and the low $\alpha_{\rm IMF}$ spectrum show a Kroupa-like \citep{Kroupa(2001)} (or even lighter) IMF, which is inconsistent with what their average $\alpha_{\rm IMF}$ indicates ($\lg\,\alpha_{\rm IMF,avr}=0$ for high $\alpha_{\rm IMF}$ sub-group, indicating a Salpeter IMF). The high $\alpha_{\rm IMF}$ spectrum shows higher $\mathrm{Na_I}$ equivalent width than the low $\alpha_{\rm IMF}$, consistent with what we see in \autoref{fig:spec_part1}. SS4 galaxies show larger difference between the high $\alpha_{\rm IMF}$ and the low $\alpha_{\rm IMF}$ spectra, where high $\alpha_{\rm IMF}$ spectrum appears to have a Salpeter-like IMF (under the VCJ model) and the low $\alpha_{\rm IMF}$ spectrum appear to have a Kroupa-like IMF, consistent with what the average $\alpha_{\rm IMF}$ indicates. For SS2 and SS3, however, we again see the opposite trend for the two spectra.

\citet{Saglia_et_al.(2002)} pointed out that the IMF shape and chemical abundances play a complex role in affecting the near-infrared features. Actually, the equivalent width of $\mathrm{Na_I}$ is not only affected by the IMF shape, but also by the abundance of sodium \citep[e.g.][]{Parikh_et_al.(2018)}. For example, \citet{Jeong2013} reported that a sodium abundance dependent on $\sigma_{\rm e}$ could also mimic an IMF variation. Here, we employ the equivalent width of $\mathrm{NaD}$, which is sensitive to the sodium abundance and thus a good indicator, to account for the influence of sodium abundance on the $\mathrm{Na_I}$-IMF relation. The feature region and the corresponding blue and red pseudo-continuum of $\mathrm{NaD}$ are defined in \citet{Trager_et_al.(1998)}. The equivalent width of $\mathrm{NaD}$ is calculated following the same steps of $\mathrm{Na_I}$ and is also corrected to $\sigma = 100\,\mathrm{km\cdot s^{-1}}$. In the bottom panels of \autoref{fig:spec_part2}, we show the distributions of $\mathrm{Na_I/NaD}$ for high and low $\alpha_{\rm IMF}$ spectra. The distributions are also obtained with a Monte Carlo-based method mentioned above. As can be seen, unlike the $\mathrm{Na_I}$-IMF, all the spectra with high $\alpha_{\rm IMF}$ show higher $\mathrm{Na_I/NaD}$ ratio than their low $\alpha_{\rm IMF}$ counterparts. This trend is even stronger for the spectra stacked with more galaxies (i.e. SS3 and SS4). It implies that, when accounting for the influence of sodium abundance, galaxies with different dynamical $\alpha_{\rm IMF}$ do show different IMF-sensitive features (i.e. the $\mathrm{Na_I}$ doublet) by stacking a number of galaxies together.

In summary, our analysis of the stellar absorption features presented in this section is not necessarily inconsistent with the variation of the dynamically-determined $\alpha_{\rm IMF}$ being due to a variation of the IMF from Chabrier-like to Salpeter-like as a function of $\sigma_{\rm e}$. However, it is undeniable that the spectra with different $\alpha_{\rm IMF}$ in \autoref{fig:spec_part1} look strikingly similar and a small trend only emerges from careful analysis of subtle effects. Moreover, we have to rely on a single IMF-sensitive spectral feature, which could be partially affected by abundance variation (although we try to asses it). The combination of these facts does not make the result from spectral features conclusive. The problem is that the models themselves predict barely detectable differences in the line depths ($\la0.3\%$) between a Kroupa/Chabrier-like and Salpeter-like IMF at fixed chemical abundance \citep[the reader is strongly encouraged to look at][fig.~1c]{van_Dokkum_et_al.(2010)}. Earlier studies from spectral features suggested significantly more bottom-heavy IMF than Salpeter, which would have been easier to detect, both with the dynamical and spectroscopic methods. However, our work excludes such extreme IMFs, at least for the mean galaxy population at fixed velocity dispersion.
 
\begin{table}
\caption{Numbers of galaxies in different sub-samples. See \autoref{sec:sample} for the selection criteria. $N_{\rm gal}$ is the number of galaxies of each sub-group. $S/N$ is the signal-to-noise ratio of the $\rm Na_I$ feature ($8182.2\,\Angstrom-8202.3\,\Angstrom$) of the stacked spectrum. $\lg\,\alpha_{\rm IMF,avr}$ is the average stellar mass excess factor of the sub-group. $\sigma_0$ is the velocity dispersion of the stacked spectrum.} \setlength{\tabcolsep}{1.8mm}
\begin{tabular}{cccccc}
\hline
\hline
Sub-sample & Sub-group & $N_{\rm gal}$ & $S/N$ & $\lg\,\alpha_{\rm IMF,avr}$ & $\sigma_0\,(\mathrm{km\cdot s^{-1}})$\\
\hline

\multirow{2}{*}{\centering SS1} & high $\alpha_{\rm IMF}$ & 21 & 2863 & 0.00 & 147.4\\
                                & low  $\alpha_{\rm IMF}$ & 21 & 2951 & -0.23 & 137.9\\
\hline
\multirow{2}{*}{\centering SS2} & high $\alpha_{\rm IMF}$ & 25 & 4289 & 0.05 & 167.0\\
                                & low  $\alpha_{\rm IMF}$ & 25 & 4011 & -0.20 & 172.2\\
\hline
\multirow{2}{*}{\centering SS3} & high $\alpha_{\rm IMF}$ & 47 & 9428 & 0.07 & 144.8\\
                                & low  $\alpha_{\rm IMF}$ & 43 & 7708 & -0.24 & 149.6\\
\hline
\multirow{2}{*}{\centering SS4} & high $\alpha_{\rm IMF}$ & 71 & 14433 & 0.05 & 170.6\\
                                & low  $\alpha_{\rm IMF}$ & 71 & 17376 & -0.23 & 138.9\\
\hline

\end{tabular}
\vspace{2mm}
\label{table:spec_samples}
\end{table}

\section{IMF discussion}
\label{sec:discussion}
We show in this discussion that there are ways to explain both (i) the lack of correlation between the $M/L$ predicted from the dynamics and from the population and (ii) the lack of correlation between IMF and metallicity from dynamics, combined with a clear correlation between IMF and metallicity from spectral features. However, explaining both facts requires some unlikely conspiracies. If we were to apply Occam's razor, a simpler explanation would be that either the dynamics, or the population, or both, do not truly measure the weight of IMF. Of course this still would require one to explain why this is the case, for which we do not have a satisfactory explanation either.

\subsection{Top or bottom heavy IMF?}
\label{sec:top_vs_bottom}
The high stellar mass excess factor can be either due to the bottom-heavy IMF, which generates more low mass stars (and hence produces a higher stellar mass-to-light ratio), or the top-heavy IMF, producing more massive stars, which are dead shortly after birth and thus contribute to the stellar mass but do not contribute to the luminosity. \citet{Barber_et_al.(2018),Barber_et_al.(2019a),Barber_et_al.(2019b)} built up a hydrodynamical cosmological simulation upon the original EAGLE Simulation \citep{Schaye_et_al.(2015)} to study the possible influence of variable IMF, by allowing the IMF to be changed according to the gas pressure. They confirmed that both top-variable and bottom-variable IMFs are able to reproduce the stellar mass excess factor-$\sigma_{\rm e}$ relation presented in \citet{Cappellari_et_al.(2013b)}. However, the analyses on gravity-sensitive spectral features (e.g. $\mathrm{Na_I}$ and $\rm FeH$) can only put constraints on the IMF slope at low mass end (typically $m<0.5\,\mathrm{M_{\odot}}$; e.g. \citealt{van_Dokkum_et_al.(2010),La_Barbera_et_al.(2013),Parikh_et_al.(2018)}). Thus, the higher dynamics-based $\alpha_{\rm IMF}$ at high $\sigma_{\rm e}$ does not necessarily mean high dwarf star enrichment (i.e. bottom-heavy IMF; see \citealt{Smith(2020)} for a review). A conspiracy between high-mass and low-mass part of the IMF (e.g. \citealt{Lyubenova2016}, and more recently, \citealt{den_Brok_et_al.(2024)}) may explain why, on the galaxy-by-galaxy basis, the stellar mass excess factors from dynamical methods do not match those from spectral feature analysis \citep{Smith(2014)}.

\subsection{Lack of IMF-metallicity relation from dynamics}
\label{sec:imf2metal_origin}
It is intuitive that the IMF shape correlates with metallicity of galaxies, as the metallicty has impact on the cooling rate of the molecular clouds, which thus influence the relative numbers of the newly-formed stars of different masses. With analyses on the gravity-sensitive spectral features (e.g. $\mathrm{Na_I}$ and $\mathrm{FeH}$) and the full spectra, the correlation between metallicity and IMF shape has been firmly detected (e.g. \citealt{Conroy_et_al.(2012),Zhou_et_al.(2019),Gu_et_al.(2022)}). However, as mentioned in the previous section, the IMF constraints so derived are only for the low mass end of the IMF shape (i.e. $m\lesssim 0.5\,\mathrm{M_{\odot}}$). Actually, the slope of the IMF at high mass end also varies with the metallicity of galaxies. For example, \citet{Marks_et_al.(2012)} found that with increasing metallicity, IMF becomes more top-light, by observing the Galactic globular clusters. \citet{Yan_et_al.(2017)} (see fig.~B1 therein; and see also \citealt{Yan_et_al.(2021)}) also found the similar decreasing trend of IMF top slope (getting steeper, i.e. more top-light) with increasing metallicity. Thus, combining the IMF studies on the low mass end \citep[e.g.][]{van_Dokkum_et_al.(2010),Parikh_et_al.(2018)} and on the high mass end \citep[e.g.][]{Marks_et_al.(2012),Yan_et_al.(2017)}, we find that the initial mass function of galaxies gets more bottom-heavy and top-light with increasing metallicity, the former of which results in higher stellar mass excess factor, while the latter results in the lower. This could qualitatively (at least partially) explain why we cannot see the correlation between the dynamics-based stellar mass excess factor and metallicity of galaxies. We note here that the metallicity dependence of IMF slope at the high mass end is somehow weak and may not be able to fully account for the absence of (dynamics-based) stellar mass excess factor-metallicity correlation. To solve this problem, we need a more detailed galaxy formation model which simulates the co-evolution of IMF and metallicity (e.g. fig.~5 of \citealt{Yan_Z_et_al.(2019)}). A hydrodynamical cosmological simulation, which enables the IMF variation may also help. 

\section{Conclusion}
\label{sec:conclusion}
This work is the 5th paper of our MaNGA DynPop (Dynamics and stellar Population) series (see \citealt{Zhu_et_al.(2023a),Zhu_et_al.(2024),Lu_et_al.(2023a),Wang_et_al.(2024)} for the other papers of this series). In this work, we study the dark matter fraction and initial mass function (IMF) variations among the galaxies in the final data release of MaNGA project \citep{MaNGA_dr17}. Specifically, we take the dynamically-determined stellar mass-to-light ratio with Jeans Anisotropic Modelling (JAM; \citealt{Cappellari2008,Cappellari2020}) from the first paper of MaNGA DynPop series \citep{Zhu_et_al.(2023a)} and the stellar mass-to-light ratio with stellar population synthesis from the second paper of MaNGA DynPop \citep{Lu_et_al.(2023a)} (assuming a Salpeter IMF), which are, by design, estimated within the same aperture, and hence are suitable for this study. In addition, we construct a new JAM dynamical model that relaxes the assumption of a constant stellar $M_{\ast}/L$, to assess the sensitivity of our results to the  possible presence of IMF gradients.

By selecting a sub-sample of galaxies with the best dynamical modelling quality, and calculating the stellar mass excess factors (i.e. the ratio between the stellar mass-to-light ratios from two different methods; see \autoref{eq:alpha} for definition), we are able to study the correlations between IMF and galaxy properties (e.g. velocity dispersion, age, metallicity, and SPS-based stellar mass-to-light ratio). Further, we also investigate the possible spectral difference between galaxies with high and low stellar mass excess factors. The main conclusions of this work are summarized below:
\begin{enumerate}
	\item The total (including stars and dark matter; obtained from JAM with an MFL model) mass-to-light ratio $\lg\,(M/L)_{\rm JAM}$ within a sphere of radius $1R_{\rm e}$ shows a parabolic correlation with the velocity dispersion of galaxies $\lg\sigma_{\rm e}$, while the stellar mass-to-light ratios, both $\lg\,(M_{\ast}/L)_{\rm JAM}$ from JAM and $\lg\,(M_{\ast}/L)_{\rm SPS}$ from SPS, prefer a linear relation (\autoref{fig:ml2sigma}). The large difference between total mass-to-light ratio and stellar mass-to-light ratio at low $\sigma_{\rm e}$ indicates a clear rise of the dark matter fraction within $1R_{\rm e}$ towards low $\sigma_{\rm e}$, which is confirmed in \autoref{fig:fdm2sigma}. We also provide an empirical relation between JAM-based dark matter fraction (within a sphere with radius being $R_{\rm e}$) and $\lg\,\sigma_{\rm e}$ (see \autoref{eq:fdm2sigma} and \autoref{table:table_eqfdm}).
	
	\item SPS-based stellar mass-to-light ratio of low $\sigma_{\rm e}$ (and low $(M_{\ast}/L)_{\rm JAM}$) galaxies is higher than JAM-based stellar mass-to-light ratio, indicating that the assumed Salpeter IMF is too heavy for low $\sigma_{\rm e}$ galaxies. On the contrary, galaxies with high $\sigma_{\rm e}$ and high $(M_{\ast}/L)_{\rm JAM}$ have higher $(M_{\ast}/L)_{\rm JAM}$ than $(M_{\ast}/L)_{\rm SPS}$, indicating that the assumed Salpeter IMF is too light for those galaxies (\autoref{fig:mlcompare}). This is consistent with the findings in \citet{Cappellari_et_al.(2012)}.
	
	\item The stellar mass excess factor, $\alpha_{\rm IMF}$ shows positive correlation with velocity dispersion, $\sigma_{\rm e}$, indicating that galaxies with $\lg\sigma_{\rm e}\la2.0$ tend to have Chabrier-like (or even lighter) IMF, while those with high $\sigma_{\rm e}$ tend to have Salpeter-like IMF. This is consistent with the previous findings in which the similar approach is adopted \citep[e.g.][]{Cappellari_et_al.(2013b),Li_et_al.(2017),Shetty_et_al.(2020)}. With different dynamical models (see \autoref{sec:jam_models} for the definition of the models), the positive correlation is still seen, but with the slope of the trend slightly changing. The mass-follows-light model (MFL), which has the fewest free parameters gives the tightest $\lg\,\alpha_{\rm IMF}-\lg\,\sigma_{\rm e}$ relation (with the scatter being 0.087), while the gNFW model shows the largest scatter ($\Delta = 0.097$; see \autoref{fig:imf2sigma_model}).
	
	\item The $\lg\,\alpha_{\rm IMF}-\lg\,\sigma_{\rm e}$ relation is sensitive to the sample selection criteria. Apart from selecting galaxies according to their morphologies (the default choice of this work), we also select galaxies based on their dust attenuation effect (i.e. $A_V<0.2$, where $A_V$ is the dust attenuation effect at $V-$band, i.e. $\lambda=5500\,\Angstrom$) and global age (i.e. $\lg\,\mathrm{Age/yr}>9.7$). We find that, although the positive correlation between $\alpha_{\rm IMF}$ and $\sigma_{\rm e}$ is still seen, the slopes of the correlation are shallower and the scatters are larger, compared to the galaxy sample selected according to their morphologies (i.e. the morphology-based sample; \autoref{fig:imf2sigma_sample}).
	
	\item By selecting a sample of early-type galaxies with low stellar mass-to-light ratio gradient (i.e. $|\gamma_{M_{\ast}/L}|<0.1$), we study the influence of the constant $M_{\ast}/L$ assumption in JAM on the $\lg\,\alpha_{\rm IMF}-\lg\,\sigma_{\rm e}$ relation. We find that the $\lg\,\alpha_{\rm IMF}-\lg\,\sigma_{\rm e}$ relation does not show obvious difference from the parent sample (i.e. the morphology-based sample; \autoref{fig:imf2sigma_sample}). Further, with a new JAM model which makes no assumption on the mass-to-light ratio (see \autoref{sec:constant_ML_test}), we investigate the effect of the possible IMF radial variation within individual galaxies on our results. We show that the inferred $\alpha_{\rm IMF}$ trend cannot be driven by radial gradients of $M_{\ast}/L$ (with fixed IMF) or IMF. We also find no evidence for the total $M/L$ increasing toward the centre within $1R_{\rm e}$ (\autoref{fig:ml_comparison_pl}, \autoref{fig:alpha2sigma_pl}, and \autoref{fig:gamma_comparison}).
	
	\item Positive correlation between $\alpha_{\rm IMF}$ and age and stellar mass-to-light ratio is observed, indicating that galaxies with older age and higher stellar mass-to-light ratio are more likely to have heavier IMF, consistent with previous studies \citep{Cappellari_et_al.(2012),McDermid_et_al.(2014)}. No obvious correlation between $\alpha_{\rm IMF}$ and metallicity is seen (\autoref{fig:imf2sp}), consistent with \citet{McDermid_et_al.(2014)}, but inconsistent with those with a spectroscopic method (e.g. \citealt{Conroy_et_al.(2012),Zhou_et_al.(2019)}). This long-standing crucial discrepancy may indicates that either the dynamical modelling-based method or the spectroscopic method or both, may not be measuring the IMF after all.
	 
	\item By stacking the spectra of galaxies with high $\alpha_{\rm IMF}$ and low $\alpha_{\rm IMF}$, we study the $\mathrm{Na_I}$ difference (which is sensitive to the IMF shape at low mass end) between the samples with different (dynamical modelling-based) IMFs. A clear result is the strikingly similarity of the stacked spectra around the $\mathrm{Na_I}$ feature. Any small difference highly depends on the sample selection criteria (\autoref{fig:spec_part1} and the top panels of \autoref{fig:spec_part2}). Our quantitative analysis of the $\mathrm{Na_I}$ feature shows that this similarity is not necessarily inconsistent with the IMF varying from Chabrier to Salpeter with $\sigma_{\rm e}$. However, this would require yet another conspiracy.
\end{enumerate}

In summary, this study used a subset of the most accurate dynamical models based on MaNGA IFU data and a consistent analysis of IMF-sensitive spectral features for the same galaxies. However, the main finding about the IMF is the absence of a clear evidence for the agreement between the two IMF indicators. This implies possible problems with either or both of them.

\section*{Acknowledgements}
We acknowledge the anonymous referee for valuable comments which improved the paper. We also acknowledge Dr. Zhiqiang Yan for helpful discussions on the metallicity-dependence of the high-mass-end slope of the initial mass function (IMF). This work is partly supported by the National Key Research and Development Program of China (No. 2018YFA0404501 to SM), by the National Natural Science Foundation of China (Grant Nos. 11821303, 11761131004, and 11761141012). This project is also partly supported by Tsinghua University Initiative Scientific Research Program ID 2019Z07L02017. We also acknowledge the science research grants from the China Manned Space Project with No. CMS-CSST-2021-A11. KZ and RL acknowledge the support of National Natural Science Foundation of China (Nos. 11988101, 11773032, 12022306), the support from the Ministry of Science and Technology of China (No. 2020SKA0110100),  the science research grants from the China Manned Space Project (Nos. CMS-CSST-2021-B01, CMS-CSST-2021-A01), CAS Project for Young Scientists in Basic Research (No. YSBR-062), and the support from K.C.Wong Education Foundation.

Funding for the Sloan Digital Sky Survey IV has been provided by the Alfred P. Sloan Foundation, the U.S. Department of Energy Office of Science, and the Participating Institutions. 

SDSS-IV acknowledges support and resources from the Centre for High Performance Computing  at the University of Utah. The SDSS website is www.sdss.org.

SDSS-IV is managed by the Astrophysical Research Consortium for the Participating Institutions of the SDSS Collaboration including the Brazilian Participation Group, the Carnegie Institution for Science, Carnegie Mellon University, Centre for Astrophysics | Harvard \& Smithsonian, the Chilean Participation Group, the French Participation Group, Instituto de Astrof\'isica de Canarias, The Johns Hopkins University, Kavli Institute for the Physics and Mathematics of the Universe (IPMU) University of Tokyo, the Korean Participation Group, Lawrence Berkeley National Laboratory, Leibniz Institut f\"ur Astrophysik Potsdam (AIP),  Max-Planck-Institut 
f\"ur Astronomie (MPIA Heidelberg), Max-Planck-Institut f\"ur Astrophysik (MPA Garching), Max-Planck-Institut f\"ur Extraterrestrische Physik (MPE), National Astronomical Observatories of China, New Mexico State University, New York University, University of Notre Dame, Observat\'ario Nacional / MCTI, The Ohio State University, Pennsylvania State University, Shanghai Astronomical Observatory, United Kingdom Participation Group, Universidad Nacional Aut\'onoma de M\'exico, University of Arizona, University of Colorado Boulder, University of Oxford, University of Portsmouth, University of Utah, University of Virginia, University of Washington, University of Wisconsin, Vanderbilt University, and Yale University.
 
\section*{Data availability}
The dynamical and structural properties of MaNGA galaxies used in this work are from the first paper of our MaNGA DynPop series \citep{Zhu_et_al.(2023a)}. The stellar population properties of these galaxies are from the second paper of this series \citep{Lu_et_al.(2023a)}. The readers can obtain all these data from the website of MaNGA DynPop (\url{https://manga-dynpop.github.io}). The catalog from the power-law JAM model is new from this paper and were included as supplementary material and also included to the DynPop catalog.

\bibliographystyle{mnras}
\bibliography{ref} 

\label{lastpage}
\end{document}